\documentclass{article}
\usepackage{arxiv} 

\usepackage[utf8]{inputenc} 
\usepackage[T1]{fontenc}    
\usepackage{hyperref}       
\usepackage{url}            
\usepackage{booktabs}       
\usepackage{amsfonts}       
\usepackage{nicefrac}       
\usepackage{microtype}      
\usepackage{amsmath, bbm, amssymb}
\usepackage{subcaption}
\usepackage{graphicx}
\usepackage{todonotes}
\usepackage{algorithm, algpseudocode}
\graphicspath{ {./images/} }
\usepackage[numbers, compress]{natbib}

\usepackage[dvipsnames]{xcolor}
\definecolor{oi_vermillion}{RGB}{213,  94,   0}
\definecolor{oi_blue}{RGB}{        0, 114, 178}
\definecolor{oi_yellow}{RGB}{    240, 228,  66}
\definecolor{oi_green}{RGB}{       0, 158, 115}

\title{Increasing Inter-Fiber Contact in  the Altendorf-Jeulin Model}

\author{
 Alex Keilmann\thanks{Corresponding author: \texttt{keilmann@rptu.de}}\\
 Department of Mathematics\\
 RPTU Kaiserslautern-Landau\\
 ORCID: 0009-0004-9793-3065
 \And 
 Claudia Redenbach\\
 Department of Mathematics\\
 RPTU Kaiserslautern-Landau\\
 ORCID: 0000-0002-8030-069X
 \And 
 Fran\c{c}ois Willot\\
 Center of Mathematical Morphology (CMM)\\
 Mines Paris Tech\\
 ORCID: 0000-0003-1544-6550
}

\begin{document}
\maketitle
\begin{abstract}

In fields such as material design or biomedicine, fiber materials play an important role. Fiber simulations, also called digital twins, provide a basis for testing and optimizing the material's physical behavior digitally. Inter-fiber contacts can influence the thermal and mechanical behavior of a fiber system; to our knowledge, however, there exist no parametric fiber models allowing for explicit modeling of the number of inter-fiber contacts. Therefore, this paper proposes an extension of the iterative force-biased fiber packing by Altendorf \& Jeulin. In this extension, we model the inter-fiber contacts explicitly and add another force to the force-biased packing to increase the number of contacts. We successfully validate the packing with respect to its parameter accuracy. Moreover, we show that the extension indeed increases the number of contacts, even exceeding theoretical values. Hence, this packing scheme has the potential to achieve higher accuracy in physical simulations. 

\end{abstract}
\section{Introduction}

Fiber materials are of great interest in areas such as material design or biomedicine due to their advantageous properties, which depend on their microstructure. Therefore, fiber simulations are used to test and optimize material behavior, also known as creating digital twins. For example, Andr\"a et~al.~\cite{andra2023ImagebasedMicrostructuralSimulation} modeled fiber insulation mats based on CT images and systematically varied parameters to study their impact on thermal conductivity. Such an approach is more sustainable than producing a variety of materials and testing their properties, as is traditionally done.

However, for digital twins to be successful, they need to model the relevant characteristics of the fiber system to ensure that they are representative of the physical behavior. On the level of a single fiber, this includes the fiber size, shape, curvature, and direction distribution. On the scale of the fiber system, the interaction between fibers can be of great interest: Whereas unrealistic fiber intersections are negligible in some applications~\cite{berhan07Modeling}, they are prohibitive in others. Fiber models allowing for such intersections are called softcore models, whereas models prohibiting them are called hardcore models.

Algorithms for simulating hardcore fiber systems can generally be categorized into two distinct approaches, the sequential approach and the collective rearrangement approach. In sequential approaches, fibers are added sequentially to the fiber system such that they do not intersect. In collective rearrangement, the fibers are generated simultaneously and may intersect in the initial configuration. In a subsequent step, their intersections are removed.

One of the first models to achieve a hardcore system is the random sequential adsorption (RSA) model. Here, elements such as straight cylinders are generated independently following a given distribution on length, radius, and/or orientation. These elements are placed iteratively into the existing system if they do not cause any intersections. Otherwise, they are resampled. However, these models allow for rather low volume fractions~\cite{pan2008Analysis3DRandom, redenbach2011StatisticalAnalysisStochastic}. There are several variations and extensions of the original RSA approach, such as allowing fibers to "grow" until they intersect other fibers~\cite{tian2015RepresentativeVolumeElement}. This allows for higher volume fractions up to 22.5\,\% at the cost of less control over the length distribution.

Sedimentation models~\cite{provatas2000FiberDepositionModels, niskanen1994PlanarRandomNetworks, venkateshan2016ModelingEffectsFiber, moghadam2019CharacterizingNonwovenMaterials} usually describe fiber systems that are obtained by fibers subsequently falling on top of each other. They can achieve higher volume fractions up to 80\,\%. However, their fiber direction distributions are roughly limited to the plane.

When collective rearrangement approaches remove intersections, they calculate the direction and amount of object displacement based on forces, which in turn are modeled on the basis of physics laws. In an early collective rearrangement approach, a fiber system is generated and, afterwards, fiber intersections are removed depending on the solution of a differential equation. It allows simulation of packings of spherocylinders~\cite{williams2003RandomPackingsSpheres} or ball chains~\cite{karayiannis2008DenseNearlyJammed}. Schneider and colleagues\cite{schneider2017SequentialAdditionMigration, schneider2022AlgorithmGeneratingMicrostructures, mehta2022SequentialAdditionMigration, lauff2024GeneratingMicrostructuresLong, lauff2025MicrostructureGenerationLong, lauff2025InfluenceFiberCurvature} transformed this approach into an optimization problem, including curved fibers. In addition, they found that the algorithm performs better when fibers are added to the system sequentially.  Venkateshan et~al.~\cite{venkateshan2016ModelingEffectsFiber} and Klar et~al.~\cite{klar19InteractingFiberStructures, borsche2017RetardedMeanFieldApproach} propose sedimentation models with collective rearrangement using differential equations.

As another subcategory of collective rearrangement approaches, force-biased approaches are inspired by molecular dynamics. Salnikov et~al.~\cite{salnikov2015EfficientReliableStochastic} propose such a model for cylinders and spheres; Bezrukov \& Stoyan~\cite{bezrukov2006SimulationStatisticalAnalysis} introduce a force-biased model for ellipsoids. Altendorf \& Jeulin~\cite{altendorf2011RandomwalkbasedStochasticModeling} extend this approach to more complex fiber shapes: They consider fibers as ball chains that are generated by a random walk. To remove intersections, repulsion forces act on overlapping balls. To maintain the fibrous structure, recovery forces act between neighboring balls of a chain. Moreover, the recovery forces preserve a prescribed fiber curvature. This approach was extended by Easwaran~\cite{easwaran2017StochasticGeometryModels} for infinite fibers and fiber bundles; Chapelle et~al.~\cite{chapelle2015GENERATIONNONOVERLAPPINGFIBER} improve the model's runtime by using spherocylinders instead of balls. Note that the molecular dynamics approach is also not limited to collective rearrangement approaches - Kumar et~al.~\cite{kumar2024MicrostructureGenerationAlgorithm} develop a fiber model based on the movement paths of charged particles, which are introduced sequentially.

In various applications, not only the (lack of) fiber intersections but also the contact between fibers is of interest. For example, inter-fiber contact areas influence the thermal conductivity of fiber systems~\cite{Gaunand07Modeling, andra2023ImagebasedMicrostructuralSimulation}. 
Friction, adhesion, and wear along contact points~\cite{yastrebov2013numerical} 
are an essential feature of entangled mechanical networks~\cite{picu2011mechanics}
such as textiles~\cite{xie2023mechanics}.
They must be taken into account to model
the significant strain rates and hysteresis phenomena
observed in many industrial compaction processes~\cite{poquillon2005experimental}.
Micromechanical models, pioneered by Wyk~\cite{van194620},
have established non-trivial
scaling laws for the effective mechanical response as
a function of the fiber density
and for the number of contact points with respect to the
fiber volume fraction~\cite{durville2005numerical}.

A common approach~\cite{faessel20053DModellingRandom, karakoc2017GeometricalSpatialEffects, Gaunand07Modeling} to model contact between fibers is the explicit transformation of fiber intersections into contact areas, thus dropping the hardcore condition. Implicitly, Altendorf \& Jeulin's force-biased packing~\cite{altendorf2011RandomwalkbasedStochasticModeling} achieves the same. However, it has not yet been thoroughly researched how to model inter-fiber contact parametrically.

For cellular fiber networks, i.e., fibers that are connected at their endpoints, inter-fiber contact is modeled more explicitly. Deogekar et~al.~\cite{deogekar2019RandomFiberNetworks} model them as stochastic Voronoi networks with a connectivity parameter, which is constant for each network. Huisman et~al.~\cite{huisman2008MonteCarloStudy} randomly change the topology using a Monte Carlo minimization scheme. In their earlier work~\cite{huisman2007ThreeDimensionalCrossLinkedFActin}, they employ a force-biased approach in which fiber endpoints within a certain distance attract each other. The network topology, especially the number of contact points, is controlled via the force strength.

In the present paper, we propose a parametric hardcore fiber model where the number of inter-fiber contacts can be increased explicitly. We extend the force-biased approach by Altendorf \& Jeulin~\cite{altendorf2011RandomwalkbasedStochasticModeling} by another force to increase the number of contacts analogous to Huisman et~al.~\cite{huisman2007ThreeDimensionalCrossLinkedFActin}. The paper is structured as follows. Section~\ref{sec:analyticalModels} reviews an analytical quantification of contact in softcore models; Section~\ref{sec:aj-model} recapitulates the Altendorf-Jeulin model~\cite{altendorf2011RandomwalkbasedStochasticModeling}. In Section~\ref{sec:extension}, we introduce our model extension, starting with algorithmic notions of contact and closeness. Runtime, feature accuracy, and the maximum contact densities achievable are evaluated in Section~\ref{sec:validation}. In Section~\ref{sec:application}, we replicate models for a real-world data set, namely a wood fiber insulation mat considered by Andr\"a et.~al.~\cite{andra2023ImagebasedMicrostructuralSimulation}, and we close with a discussion and conclusion in Section~\ref{sec:conclusion}. 

\section{State of the Art}
\subsection{Analytical Description for Fiber Contact}
\label{sec:analyticalModels}

Consider a stationary fiber system in a compact window $W \subset \mathbb{R}^3$ with intensity (mean number of fibers per unit volume) $\lambda_F$. 
We assume that the fibre orientations follow a distribution with density $\psi$. 
We assume that the fibers have constant
radius $r$ and variable length $\ell$ and denote by
$\bar{\ell}$ the expected value of the fiber length distribution. 
We denote by $\lambda_c$ the mean number of inter-fiber contact points per unit volume, i.e., the intensity of inter-fiber contacts.
The expected number of contacts per fiber is $\lambda_{cF}=\lambda_c/\lambda_F$.

Toll~\cite{toll1993NoteTubeModel} 
provides the following estimate
for the intersection intensity:
\begin{subequations}
\begin{eqnarray}\label{eq:toll}
    \lambda_{cF} &\approx& 4\lambda_F r \bar{\ell}\left[  \bar{\ell} f_\psi + \pi r (g_\psi+1)\right],
    \qquad
    \lambda_{c}\approx 4 \lambda_F^2 r \bar{\ell}\left[ \bar{\ell} f_\psi + \pi r (g_\psi + 1)\right],\\
    f_\psi &=& \int_{S^2} \int_{S^2} |p \times p'|\psi(p')\psi(p)\,dp' \,dp,
    \qquad
    g_\psi = \int_{S^2} \int_{S^2}|p \cdot p'|\psi(p')\psi(p)\,dp' \,dp,
\end{eqnarray}
\end{subequations}
where $\times$ and $\cdot$ denote the vector and scalar products in $\mathbb{R}^3$.
We refer to~\cite{toll1998PackingMechanicsFiber}
for formulas obtained in special cases corresponding to an isotropic or transversely-isotropic fiber-distribution
and to ~\cite{komori1977NumbersFiberFiberContacts}
for strongly elongated fibers.
In effect, the approximations
in~\eqref{eq:toll} 
for $\lambda_{cF}$
and $\lambda_{c}$ 
can be rigorously reinterpreted as the number of \emph{intersections} per unit volume and fiber, respectively, in a Poisson-Boolean random set of cylinders with mean length $\bar{\ell}$ and orientation distribution $\psi$
(see Appendix~\ref{sec:Poiss}).
In the following, the estimates~\eqref{eq:toll} will be compared with numerical simulations.

\subsection{The Altendorf-Jeulin Model}
The fiber model by Altendorf \& Jeulin~\cite{altendorf2011RandomwalkbasedStochasticModeling} specifies a force-biased packing of fibers that are modeled as ball chains. Its starting configuration of fibers is obtained by simulating the fibers independently by a random walk in a closed cubic window $W = [0, w_1] \times [0, w_2] \times [0, w_3]$ with periodic boundary conditions. 
To preserve the periodic boundary conditions, we use the periodic equivalent $d_P$ of the Euclidean distance. For $x, x' \in \mathbb{R}^3$ and $i \in \{ 1, 2, 3\}$, let 
\begin{equation}
    \Delta_i (x, x') = \left\{\begin{array}{ll}
        x'_i - x_i - w_i, & \text{for } x'_i - x_i \geq \frac{w_i}{2} \\
        x'_i - x_i + w_i, & \text{for } x'_i - x_i \leq  \frac{-w_i}{2}\\
        x_i' - x_i, & \text{otherwise}
        \end{array}\right .
\end{equation}
Then, $d_P(x, x') = \| \Delta(x, x')\|_2$ is the periodic distance function and
\begin{equation}
    v_P(x, x') = \frac{\Delta(x, x')}{\| \Delta(x, x')\|_2}
\end{equation}
the corresponding direction of the vector between two points.

After the random walk, the fibers may intersect but follow given distributions regarding model features such as fiber size and direction. The force-biased packing removes the intersections, while keeping the model features approximately intact. Note that this packing scheme can also be applied to fiber systems that were generated differently, e.g., with a different random walk~\cite{gaiselmann2013Stochastic3DModeling}. In the following, we will recap the main features of the algorithm.

\label{sec:aj-model}
\subsubsection{Formalizing the Fiber System}
In the initial fiber system, a single fiber $\mathcal{F} = \{b_1, ..., b_l\}$ is modeled as a chain of $l$ overlapping balls $b_i = (x_i, \mu_i, r), i = 1, ..., l$, of radius $r$ and center $x_i \in \mathbb{R}^3$. In addition, we record the current direction $\mu_i \in S^2$ of the random walk. More precisely, $\mu_i = x_i - x_{i-1}$ is the direction between the last ball and the current ball for $i > 1$; for the start of the random walk in $i = 1$, $\mu_i$ indicates the main direction of the whole fiber.
In this paper, the radius $r$ of the balls and the chain length $l$ are constant for the sake of simplicity, but they can be randomized (see also ~\cite{altendorf2011RandomwalkbasedStochasticModeling}). Note that the fiber length $\ell$ will typically differ from the chain length $l$.

The point $x_1 \in \mathbb{R}^3$ denotes the starting point of a fiber and can be sampled from a uniform distribution on $W$. The fiber's main direction $\mu_1 \in S^2$ is sampled from a Schladitz distribution~\cite{schladitz2006DesignAcousticTrim} with anisotropy parameter $\beta > 0$. Its density w.r.t. polar coordinates $(\theta, \phi)$ is given by
\begin{equation}\label{eq:betapdf}
    p_S(\theta, \phi | \beta) = \frac{1}{4\pi}\frac{\beta \sin \theta}{(1 + (\beta^2 - 1)\cos^2 \theta)^\frac{3}{2}}, ~~~ \theta \in [0, \pi), \phi \in [0, 2\pi).
\end{equation}
For $\beta$ close to 0, this distribution generates directions that are mostly aligned with the z-direction; $\beta = 1$ corresponds to the uniform distribution on the unit sphere; $\beta \gg 1$ results in girdle distributions with directions that are concentrated around the equator.

The fiber bending is modeled based on a multivariate von Mises-Fisher (mvMF) distribution with parameters $\kappa_1, \kappa_2 > 0$. A univariate von Mises-Fisher distribution~\cite{fisher1987StatisticalAnalysisSpherical} has the density
\begin{equation}
    p_{M}(x | \mu, \kappa) = \frac{\kappa}{4\pi \sinh \kappa} \exp \left( \kappa \mu^T x \right), \quad x\in S^2
\end{equation}
with anisotropy parameter $\kappa \geq 0$ and direction $\mu \in S^2$. For higher $\kappa$, the fibers are more concentrated around the direction $\mu$. The multivariate version $\text{mvMF}(\mu', \kappa_1, \mu'', \kappa_2)$ of the von Mises-Fisher distribution uses $\kappa = |\kappa_1 \mu' + \kappa_2 \mu''|$ and $\mu = \frac{\kappa_1 \mu' + \kappa_2 \mu''}{\kappa}$.

Having defined the mvMF distribution, we can return to the definition of the random walk: The direction $\mu_{i+1}$ is generated following the $\text{mvMF}(\mu_1, \kappa_1, \mu_i, \kappa_2)$-distribution. Hence, $\kappa_1$ should indicate the reliability of the random walk to the main fiber direction $\mu_1$ and $\kappa_2$ the reliability to the last direction $\mu_i$. 
This defines the new position $x_{i+1} = x_i + \frac{r}{2}\mu_{i+1}$. At the end of the random walk, the fiber is rotated such that its main fiber direction $\Bar{\mu} := \frac{x_l - x_1}{\|x_l - x_1\|_2}$ equals $\mu_1$~\cite{altendorf2011RandomwalkbasedStochasticModeling}. In the implementation, the fibers are prepacked using a modified version of RSA~\cite{pan2008Analysis3DRandom, altendorf20113DMorphologicalAnalysis}: Iteratively, the starting point of each fiber is randomly assigned for a given number of trials. The fiber is eventually placed at the position of minimal overlap compared to the other trials. Other than RSA, this version does not resample fibers; instead, it uses the fibers generated previously. In addition, overlap is not avoided completely, just reduced. This later improves runtime, as fewer intersections must be removed during force-biased packing. 

The total fiber system is specified by a graph $\mathcal{S} = (\mathcal{B}_S, \mathcal{F}_S)$, where the node set
\begin{equation}
    \mathcal{B}_S = \{b_{1, 1}, b_{2, 1}, ..., b_{l, 1}, b_{1, 2}, ..., b_{l, n}\}
\end{equation}
contains the balls of the $n$ fibers, and the undirected edge set
\begin{equation}
    \mathcal{F}_S = \left\{\{b_{i, j}, b_{i + 1, j}\}| i \in \{1, ..., l - 1\}, j \in {1, ..., n} \right\}
\end{equation}
contains the connections between the balls as specified by the random walk. As a consequence, the connected components of $\mathcal{S}$ are equivalent to the fibers. We indicate that balls $b, c \in \mathcal{B}_S$ belong to the same fiber by writing $b \sim_F c$.

\subsubsection{Force-biased Packing}

\begin{figure}[!ht]
    \centering
    \begin{subfigure}{0.45\textwidth}
        \includegraphics[width=0.85\textwidth]{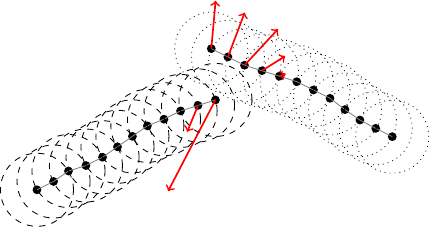}
        \caption{Sketch of repulsion forces acting on intersecting fibers.} \label{fig:repulsionSketch}
    \end{subfigure}\hspace{10pt}
    \begin{subfigure}{0.45\textwidth}
        \includegraphics[width=0.85\textwidth]{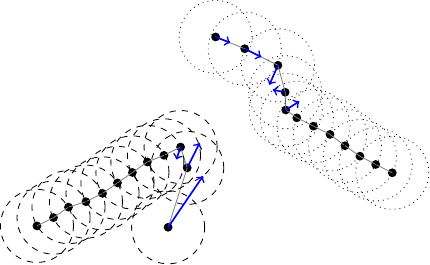}
        \caption{Sketch of recovery forces acting on corrupted fibers.}\label{fig:recoverySketch}
    \end{subfigure}
    \caption{Sketches of forces in the Altendorf-Jeulin model.}
\end{figure}

To remove the initial fiber intersections, the force-biased packing by Altendorf \& Jeulin~\cite{altendorf2011RandomwalkbasedStochasticModeling} applies repulsion forces to overlapping balls, see Fig.~\ref{fig:repulsionSketch}. Afterwards, recovery forces restore each fiber's connectedness and features, see Fig.~\ref{fig:recoverySketch}. This process is repeated iteratively until the forces are small enough.

The repulsion force acts on the intersecting balls and moves them apart. An exception pose balls that are "close" within the same fiber as they are designed to overlap: Due to the distance of $\frac{r}{2}$ between subsequent positions in the ball chain, balls inevitably intersect if they are connected by a path consisting of $\leq 4$ edges. To allow some room for fiber curvature, Altendorf \& Jeulin suggested excluding ball pairs from the repulsion force that are connected by a path of length $\leq 5$. For brevity, $b \sim_{5} c$ implies that there exists a path of length $\leq 5$ between $b$ and $c$, whereas $b \nsim_5 c$ denotes that such a path does not exist.

In the hardcore case, the strength of the repulsion force $F_r$ depends on the amount of intersection of the balls via
\begin{equation}
    I (b, c) = \max\left(0, 2r - d_P(x_b, x_c)\right).
\end{equation}
Let the softcore ratio $\tau \in [0, 1]$ describe the amount of intersection allowed, so two ball centers must have at least the distance $2(1- \tau) r$. Thus, $\tau=0$ models the hardcore case, while $tau=1$ allows for arbitrary overlap. Then, $I$ generalizes to 
\begin{equation}
    I_\tau (b, c) = \max\left(0, 2(1- \tau) r - d_P(x_b, x_c)\right).
\end{equation}
Its direction is $v(b, c) = v_P(x_b, x_c)$, yielding
\begin{equation}
    F_r(b, c) = \mathbbm{1}_{b \nsim_5 c} \frac{I_\tau (b, c)}{2}v(b, c).
\end{equation}

\begin{figure}[!ht]
    \centering
    \includegraphics[width=0.6\textwidth]{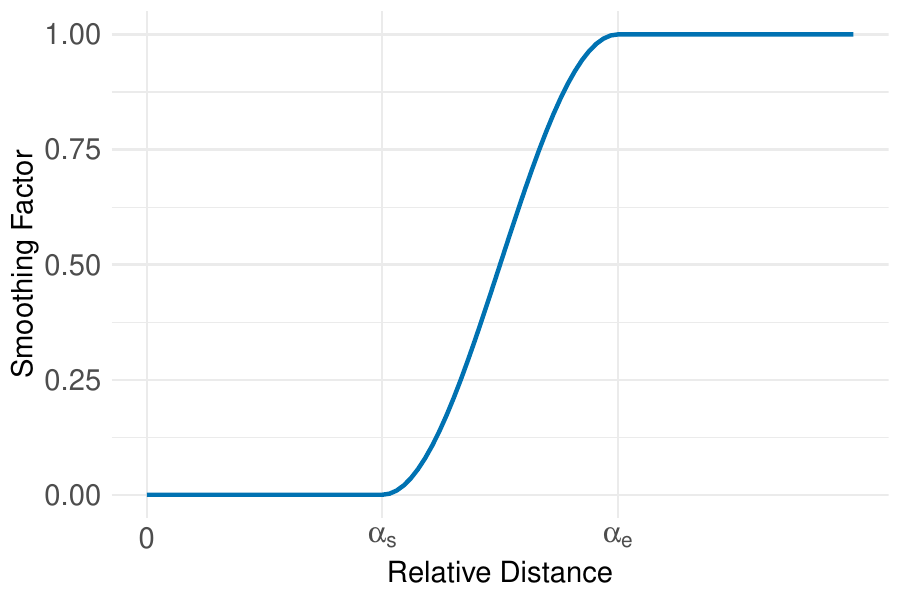}
    \caption{Plot of the smoothing factor for $0 < \alpha_s < \alpha_e$.}
    \label{fig:smoothingFactor}
\end{figure}

There are two recovery forces, which only act between neighboring balls within a single fiber: The spring force $F_s$ keeps the ball spacing within the chains intact, whereas the angle force $F_a$ preserves the fibers' bending. The spring force's strength between balls $b, c \in \mathcal{B}_S$ with $\{b,c\} \in \mathcal{F}_S$ depends on their distance, or, more precisely, on the deviation from their initial distance $\frac{r}{2}$:
\begin{equation}
    \alpha_d(b, c) = \frac{r}{2} - |x_b - x_c|.
\end{equation}
Due to the fibers' chainlike structure, the spring force easily causes ripple effects. In order to reduce them, it further incorporates a smoothing factor that depends on the function $\alpha_d$, namely

\begin{equation}
    f_{\alpha_s, \alpha_e}(\alpha) := \left\{\begin{array}{ll}
        0, & \text{for } \alpha < \alpha_s\\
        \frac{1}{2} - \frac{1}{2}\cos\left(\frac{|\alpha| - \alpha_s}{\alpha_e - \alpha_s}\pi\right), & \text{for } \alpha_s\leq \alpha\leq \alpha_e\\
        1, & \text{for } \alpha_e < \alpha
        \end{array}\right .
\end{equation}
for $\alpha_s \alpha_e$, see Fig.~\ref{fig:smoothingFactor}. For the special case of $\alpha_s = \alpha_e = 0$, we use $f_{0,0}(\alpha) := \mathbbm{1}_{\mathbb{R}^+}(\alpha)$.

The spring force is then
\begin{equation}
    F_{s}(b, c) = \mathbbm{1}_{\mathcal{F}_S}(b, c)f_{\alpha_s, \alpha_e}\left( 2\frac{|\alpha_d(b, c)|}{r}\right) \alpha_d(b, c) v(b, c).
\end{equation}
Choose $0 < \alpha_s < \alpha_e < 1$ as the argument of $f_{\alpha_s, \alpha_e}$ is a relative distance here
Due to the smoothing factor, the force is set to $0$ if $\alpha_d(b, c) < \frac{r \alpha_s}{2}$, which keeps the ripple effects of the force on the system small.

The angle force $F_a(b)$ is dependent on the required displacement of balls to preserve the fibers' local curvature. Consequently, it depends on the balls' direct neighbors, just like the spring force. As its definition is rather technical, we refer the reader to the original article~\cite{altendorf2011RandomwalkbasedStochasticModeling}.

Taken all together, the total force acting on a single ball $b \ \in \mathcal{B}_S$ is
\begin{equation}
    F_\text{total}(b) = \sum_{\substack{c \ \in \mathcal{B}_S\\ c \neq b}} F_r(b, c) + \rho\left(F_a(b) + \sum_{\{b, c\} \in \mathcal{F}_S} F_s(b, c)\right),
\end{equation}
where $\rho \in [0, 1]$ modulates the composition of repulsion and recovery forces. All parameter choices in this paper are described and justified in Section~\ref{sec:implementation}.

\section{Extension to a Contact Packing Scheme}
\label{sec:extension}
To incorporate contact in the fiber packing, we extend Altendorf \& Jeulin's~\cite{altendorf2011RandomwalkbasedStochasticModeling} fiber packing by an additional force, the contact force $F_c$. This force will be applied to balls that are close enough to establish contact. 
Before properly defining $F_c$, we need to specify what it means that two fibers are in contact.

\subsection{Algorithmic Notions of Contact and Closeness}
\label{sec:contactNotion}

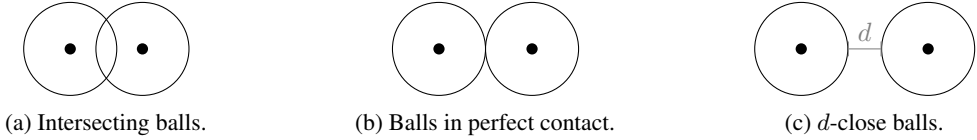
\begin{figure}[!ht]
    \begin{subfigure}{0.3\textwidth}
        \centering
        \begin{tikzpicture}[scale=0.8]
            \node[circle,fill,inner sep=1.5pt] (b1) at (0, 0) {};
            \node[circle,fill,inner sep=1.5pt] (b2) at (1.2, 0) {};
    
            \draw (b1.center) circle (22pt);
            \draw (b2.center) circle (22pt);   
        \end{tikzpicture}
        \caption{Intersecting balls.}
    \end{subfigure}
    \begin{subfigure}{0.3\textwidth}
        \centering
        \begin{tikzpicture}[scale=0.8]
            \node[circle,fill,inner sep=1.5pt] (b1) at (0, 0) {};
            \node[circle,fill,inner sep=1.5pt] (b2) at (44pt, 0) {};
    
            \draw (b1.center) circle (22pt);
            \draw (b2.center) circle (22pt);   
        \end{tikzpicture}
        \caption{Balls in perfect contact.}
    \end{subfigure}
    \begin{subfigure}{0.3\textwidth}
        \centering
        \begin{tikzpicture}[scale=0.8]
            \node[circle,fill,inner sep=1.5pt] (b1) at (0, 0) {};
            \node[circle,fill,inner sep=1.5pt] (b2) at (60pt, 0) {};
            \draw (b1.center) circle (22pt);
            \draw (b2.center) circle (22pt);   
            \draw[|-|, gray] (22pt, 0) -- (38pt, 0);
            \node[gray] at (30pt, 8pt) {$d$};
        \end{tikzpicture}
        \caption{$d$-close balls.}
    \end{subfigure}
    \caption{Sketch for the degrees of closeness on the level of balls.}
    \label{fig:contactBalls}
\end{figure}

Let the balls $b, c \in \mathcal{B}_S$ belong to different fibers, i.e., $b \nsim_F c$, then
\begin{itemize}
    \item $b$ and $c$ intersect if $|x_b - x_c| < 2r$.
    \item $b$ and $c$ are in (perfect) contact if $|x_b - x_c| = 2r$.
    \item $b$ and $c$ are $d$-close for $d \geq 0$ if $|x_b - x_c| \leq 2r + d$, denoted as $b \overset{d}{\sim} c$.
\end{itemize}
These relations are depicted in Fig.~\ref{fig:contactBalls}.

\begin{figure}[ht!]
    \centering
    \includegraphics[width=0.4\textwidth]{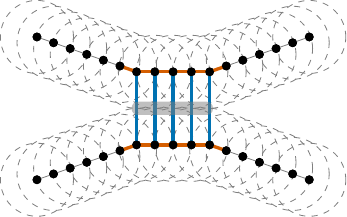}
    \caption{Sketch for the contact areas between fibers. \textcolor{oi_blue}{Blue} indicates the contact between balls as described by $\mathcal{E}_0$. \textcolor{oi_vermillion}{Red} indicates the ball chain sections creating the contact as described by $\Bar{\mathcal{E}}_0 \backslash \mathcal{E}_0$. The \textcolor{gray}{gray} area indicates the intuitive contact area.}
    \label{fig:contactFibers}
\end{figure}

Based on this contact notion, we can define a graph $\mathcal{C}_{d} = (\mathcal{B}_S, \mathcal{E}_{d})$ that describes the $d$-closeness between the fibers with the undirected edge set 
\begin{equation}
    \mathcal{E}_{d} = \left\{\{b, c\}| b, c \in \mathcal{B}_S, b \nsim_F c, \text{ and } b \overset{d}{\sim} c \right\}.
\end{equation}
In Fig.~\ref{fig:contactFibers}, the \textcolor{oi_blue}{blue lines} depict the set $\mathcal{E}_{0}$, that is, ball pairs in perfect contact. 

Apart from the contact on the level of balls, we are also interested in the contact on the level of fibers. As shown in Fig.~\ref{fig:contactFibers}, one contact area between fibers may consist of multiple ball pairs that are in contact. Using only the contact notion on the level of balls would therefore be counterintuitive as it confounds the number of ball pairs with the number of fibers. Moreover, it is dependent on the balls' size, and one ball may be in contact with more than one other ball. As a solution, we understand the contact area between two fibers as the ball chain sections where their balls are in contact. 
Since the ball chains are described by the edges in $\mathcal{F}_S$, it seems natural to extend $\mathcal{E}_{d}$ by the balls' incident edges w.r.t. $\mathcal{F}_S$. In Fig.~\ref{fig:contactFibers}, they are depicted by \textcolor{oi_vermillion}{red} lines.
To formalize this concept, we extend the edge set $\mathcal{E}_{d}$ as follows:
\begin{equation}
    \Bar{\mathcal{E}}_{d} =  \mathcal{E}_{d} \cup \left\{\{b, c\} \in \mathcal{F}_S|\exists a \in \mathcal{B}_S \text{ s.t. } \{b, a\} \in \mathcal{E}_{d} \lor \{a, c\} \in \mathcal{E}_{d}\right\}.
\end{equation}
A connected component of the graph $\Bar{\mathcal{E}}_{d}$ may contain more than two fibers; these we call \textit{clots} to discriminate them from (pairwise) contact areas between two fibers. By counting the connected components, we will later report the number of contacts per unit volume as estimator for the contact intensity, as well as the number of contacts and the number of clots per fiber as estimators of the contact and clot density, respectively. This way, we close the loop to analytical quantification for contact, see Section~\ref{sec:analyticalModels}.

Now that we have introduced a notion to quantify contact and closeness in the Altendorf-Jeulin model, we can turn to creating more contact. To do so, we identify suitable balls, to which we later apply the contact force. 
In the style of Huisman et~al.~\cite{huisman2007ThreeDimensionalCrossLinkedFActin}, we identify such balls depending on their closeness: The balls $b, c \in \mathcal{B}_S, b \nsim_F c$, are \emph{contact candidates} w.r.t. the \emph{interaction distance} $\varepsilon > 0$ if $b \overset{\varepsilon}{\sim} c$. The set of contact candidates w.r.t. the interaction distance $\varepsilon$ is then $\mathcal{E}_\varepsilon$. We deliberately include balls that are already in contact as contact candidates since these are supposed to stay in contact.

\begin{figure}[ht!]
    \centering
    \includegraphics[width=0.4\textwidth]{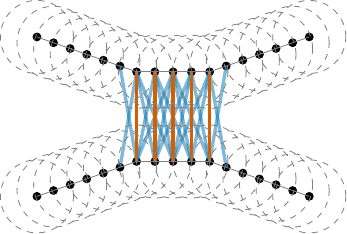}
    \caption{Sketch for the contact candidates (indicated by \textcolor{oi_blue}{blue} lines) and shortlist (indicated by \textcolor{oi_vermillion}{red} lines) for the interaction distance $\varepsilon = \frac{3}{4}r$.}
    \label{fig:shortlist}
\end{figure}

However, applying the contact force to all contact candidates in $\mathcal{E}_{\varepsilon}$ may be computationally expensive for a large $\varepsilon$. The forces might even be conflicting (see Fig.~\ref{fig:shortlist}). Therefore, we reduce the contact candidates to a "shortlist" $S_\varepsilon$, such that every ball belongs to at most one candidate pair per fiber pair. 
To achieve faster convergence, we keep the edge in the shortlist that has the minimal distance compared to incident edges. To formalize this, we define the set of incident edges
\begin{flalign}
    \mathcal{I}_{\varepsilon}(\{b, c\}) = \{\{d, e\} \in \mathcal{E}_{\varepsilon} | e\sim_F c\} \cup \{\{c, e\} \in \mathcal{E}_{\varepsilon} | e\sim_F b\}.
\end{flalign}
This set contains all edges in $\mathcal{E}_{\varepsilon}$ that belong to the node $b$ or the node $c$. If there exists an edge $\{b, c\} \in \mathcal{E}_{\varepsilon}$, then $\mathcal{I}_{\varepsilon}(\{b, c\})$ consists of all its incident edges.
We define the shortlist as
\begin{equation}
    \mathcal{S}_{\varepsilon} = \left\{\{b, c\} \in \mathcal{E}_{\varepsilon}| \{b, c\} = \underset{\{d, e\} \in \mathcal{I}_{\varepsilon}(\{b, c\})}{\text{argmin}} \| d - e\|\right\}.
\end{equation}

\subsection{Contact Force}
We extend Altendorf \& Jeulin's~\cite{altendorf2011RandomwalkbasedStochasticModeling} fiber packing by one more force, namely the \emph{contact force} $F_c$. It exists between two shortlisted contact candidates $\{b, c\} \in \mathcal{S}_{\varepsilon}$. We aim to move the balls such that they are in perfect contact, i.e.
\begin{equation}
     |x_b - x_c| = 2r,
\end{equation}
so the force depends on the difference between the ideal and actual distance. Further, there is no force necessary when the balls are already intersecting. With this, we define the \emph{effective distance}
\begin{equation}
    d_E(b, c) := \max(0, d_P(x_b, x_c) - 2r).
\end{equation}
Note that $d_E$ differs from $I$, which we introduced for the repulsion force in Section~\ref{sec:aj-model}, by exchanging subtrahent and minuend, which makes them complementary: Whereas the repulsion force pushes intersecting balls apart until they are at most in contact, the contact force draws balls together until they are at least in contact.

In analogy to the repulsion force, the contact force has the direction $v(c, b)$. Hence, we propose the contact force
\begin{equation}
    F_c(b, c) = \frac{1}{2} d_E(b, c) v(c, b)
\end{equation}
for $\{b, c\} \in \mathcal{S}_{\varepsilon}$.

In some applications, such as nanotube composites, fibers start to interact before being in physical contact~\cite{Benoit_2002}. In this case, it makes sense to consider $d_c$-close balls as being in contact for a given contact distance $d_c > 0$. Consequently, the balls $b, c \in \mathcal{B}_S, b \nsim_F c$ are contact candidates w.r.t. the interaction distance $\varepsilon > 0$ if $|x_b - x_c| \leq 2r + d_c + \varepsilon$. One can reason similarly when discretizing the fiber system, e.g. for voxelized images. There, depicting distances below the voxel size is usually impossible.

For a contact distance $d_c > 0$, we have additional leeway for balls to be considered in contact without intersection. It seems natural to use $\max(0, d_P(x_b, x_c) - (2r + d_c))$ as the effective distance, thus only aiming for $d_c$-closeness instead of perfect contact. However, this also lowers the force strength and may consequentially lead to slower convergence. Therefore, we incorporate the smoothing factor $f_{0, d_c}$ as introduced for the recovery force in Section~\ref{sec:aj-model}.
In total, we generalize the contact force to
\begin{equation}
    F_c(b, c) = \frac{f_{0, d_c}\left(d_E(b, c)\right)}{2} d_E(b, c) v(c, b)
\end{equation}
for $(b, c) \in \mathcal{S}_{\varepsilon}$.

\subsection{Synopsis of the Contact Packing Scheme}
In our contact packing scheme, we first run the regular force-biased packing by Altendorf \& Jeulin~\cite{altendorf2011RandomwalkbasedStochasticModeling}. This way, the fiber system is already (nearly) free of intersections and thus possesses lower inherent forces. 
If this system does not have the desired amount of inter-fiber contacts yet, we increase the contact by applying the packing incorporating the contact force. We explain this modified packing scheme in the following. A pseudocode is provided in Algorithm~\ref{alg:ContactPackingScheme}.

In the first step of our contact-modified fiber packing, we identify contact candidates $\mathcal{E}_{\varepsilon}$ for a given interaction distance $\varepsilon > 0$, which we then reduce to the shortlist $S_{\varepsilon}$, such that every ball belongs to only one candidate pair per fiber pair, see Section~\ref{sec:contactNotion}. In this configuration, we iteratively apply the force
\begin{equation}
    F_\text{total, c}(b) = \sum_{\substack{c \ \in \mathcal{B}_S\\ c \neq b}} F_r(b, c) + \rho_R\left(F_a(b) + \sum_{\{b, c\} \in \mathcal{F}_S} F_s(b, c)\right) + \rho_C \sum_{\{b, c\} \in \mathcal{S}_{\varepsilon}}F_c(b, c),
\end{equation}

to every ball $b \in \mathcal{F}_S$ until convergence. The stop criterion is specified in Section~\ref{sec:implementation}. Note that $\rho_R, \rho_C \in [0,1]$ are factors modulating the composition of forces analogously to $\rho$ in Section~\ref{sec:aj-model}.

Some shortlisted candidates $(b, c) \in \mathcal{S}_{\varepsilon}$ can cause conflicting forces in the fiber system even after fiber packing. This can, for example, happen in denser fiber systems when the contact forces in different contact components "pull" fibers in directions that conflict unresolvably with their repulsion and recovery forces. Note that a similar phenomenon has already been observed by Altendorf \& Jeulin~\cite{altendorf2011RandomwalkbasedStochasticModeling}, where they compare it with the jammed state~\cite{karayiannis2008DenseNearlyJammed}. We remove such shortlisted candidates and apply the force again iteratively until termination. If this does not achieve the desired amount of contact, we can repeat the procedure with an increased interaction distance $\varepsilon$. 
Shortlisted candidates are classified as causing conflicting forces if for $a = b$ or $a = c$ the force length $\|F_\text{total, c}(b)\|$
exceeds a specified threshold $t > 0$.

\begin{algorithm}
\caption{Contact Packing Scheme}\label{alg:ContactPackingScheme}
\begin{algorithmic}
\Require 
$d_c \geq 0$, interaction distance $\varepsilon$
\State generate a starting configuration of fibers
\State run the Altendorf-Jeulin fiber packing with $F_{\text{total}}$
    \State calculate the contact candidates $\mathcal{E}_{\varepsilon}$
    \State calculate the shortlist $\mathcal{S}_{\varepsilon}$
    \While{!(stop criterion)}
        \State run fiber packing with $F_{\text{total, mod}}$
    \EndWhile
    \If{some candidates in $S_{\varepsilon}$ cause unresolvable forces}
        \State remove candidates from $S_{\varepsilon}$
        \While{!(stop criterion)}
            \State run fiber packing with $F_{\text{total, c}}$
        \EndWhile
    \EndIf
\end{algorithmic}
\end{algorithm}

\subsection{Implementational Details}
\label{sec:implementation}
The original fiber packing by Altendorf \& Jeulin~\cite{altendorf2011RandomwalkbasedStochasticModeling} stops when the sum of total forces $\left\|\sum_{b \in \mathcal{B}_S}F_{\text{total}}(b)\right\|$ falls below a certain threshold. For the contact-modified fiber packing, however, we decided to stop the packing when the sum of total forces decreases by less than a specified relative amount. This has the advantage of being independent of parameters like the number of fibers or window size, as is the case in the implementation by Altendorf \& Jeulin~\cite{altendorf2011RandomwalkbasedStochasticModeling}. We chose a limit of $0.001\%$ in the present paper. To ensure termination, we stop the packing when the number of iterations exceeds 1\,000.

As a reminder, one iteration of fiber packing classicly encompasses the calculation of forces on the fiber system and, subsequently, the corresponding repositioning of balls. In this paper, we choose $t = 0.1$ to remove shortlisted candidates that cause unresolvable forces in the fiber system, as it provides a trade-off between accuracy of the fibers' curvature by removing unresolvable forces on the one hand and achieving high contact on the other hand. Alternatively, the interaction distance could be chosen higher to achieve the same amount of contact, but this would cause a higher runtime in return.

For the realizations in this paper, we use $\rho = \rho_R = \rho_C = 0.5$, indicating equal strength for both recovery and contact forces. For the spring force, we use $\alpha_s = 0.05$ and $\alpha_e = 0.1$. Unless indicated otherwise, the softcore ratio is $\tau = 0$ and the contact distance $d_c = 0$. This conforms to the parameters used by Altendorf \& Jeulin~\cite{altendorf2011RandomwalkbasedStochasticModeling}, thus allowing for direct comparison.

To detect contact candidates and intersecting balls, we follow Altendorf \& Jeulin~\cite{altendorf20113DMorphologicalAnalysis} and use the particularly fast implementation by Mosćiński et~al.~\cite{moscinski1989ForceBiasedAlgorithmIrregular}. They divide the window into subwindows with a side length $(s_x, s_y, s_z)$. To find contact or intersection candidates faster, every subwindow is given a list of balls. Notably, a subwindow's list contains both the balls, whose center is in the subwindow, and the balls, whose center is in the subwindow's neighboring subwindows. This way, only balls in a subwindow's list need to be checked for contact candidacy or overlap. For a maximal interaction distance of $\varepsilon_{\text{max}}$, we use a subwindow size of at least $\max(2.5r, (2 + \varepsilon_{\text{max}})r)$. 
We calculate the connected components of $(\mathcal{B}_S, \Bar{\mathcal{E}}_{\varepsilon})$ using UnionFind~\cite{sedgewick2011algorithms}.

\section{Experimental Validation}
\label{sec:validation}

In this section, we examine the performance of our packing scheme regarding the features of the fiber system when increasing contact. We implemented the scheme in C++ using and extending the library MAVIlib~\cite{mavi}, which Fraunhofer~ITWM develops and maintains. For further speed-up, we parallelized the calculation of forces using OpenMP. Experiments were run on 
the ITWM Beehive Cluster using an Intel(R) Xeon(R) CPU E5-2670 0 @ 2.60GHz, 125 GiB RAM, and the GNU compiler GCC 11.5.

\subsection{Experimental Setup}
\label{sec:setup}
As discussed in Section~\ref{sec:analyticalModels}, the intensity of inter-fiber contacts depends on the fibers' length, radius, intensity, and direction distribution. Therefore, we vary these parameters in our experimental setup as well. For image sizes of 64 to 640 voxels per edge, we study
\begin{itemize}
    \item short, medium, and long fibers, namely fibers with an aspect ratio $a := \frac{\ell}{2r}$ of $a = 10, 30, 50$; the radius stays constant at $r = 2$.
    \item low, medium, and high volume fractions, namely $V_V = 10\%, 20\%, 30\%$. Note that fiber systems with even higher intensities will be packed quite densely already; in this case, increasing the contact density is less meaningful.
    \item the direction distribution from aligned over the uniform distribution to girdle distributions, namely we use the Schladitz distribution with $\beta = 0.1, 0.5, 1.0, 2.0, 3.0$.
\end{itemize}
As a packing parameter, we choose the interaction distances $\varepsilon = 0.1, 0.2, ..., 0.5, 1.0$.

In this setup, we use a medium curvature of $\kappa_1 = 10, \kappa_2 = 100$, the softshell ratio $\tau = 0$, and the contact distance $d_c = 0$ as default parameters. To study the scheme's accuracy when varying these parameters, we further study the cases of low curvature, namely $\kappa_1, \kappa_2 = 100$, and high curvature, namely $\kappa_1, \kappa_2 = 10$, for medium aspect ratio and medium volume fraction. Note that varying these parameters for all parameter combinations discussed above would result in three times the runtime and computing resources, which is not merely tedious but also unsustainable given the necessary energy consumption.
All used parameter combinations can be found in Table~\ref{table:parameterCombinations} in the Appendix.

\subsection{Representative Volume Element and Runtime}
\label{sec:RVE}
The representative volume element (RVE) of a microstructure is the volume $V$ that is considered large enough to be statistically representative. Note that the RVE depends not only on the microstructure, but also on the investigated property $Z$. This means, for example, that the RVE for the volume fraction and the thermal conductivity can differ for the same microstructure. Moreover, it is not always computationally feasible to generate a microstructure that is large enough. Therefore, the concept is generalized to a combination of the size and number of realizations to be representative of the underlying microstructure. This way, it can be calculated, for example, how many small realizations are necessary when high realizations are computationally prohibitive. The following computations are based on the assumption of ergodicity~\cite{kanit2003DeterminationSizeRepresentative}.

In the present work, we are interested in the properties that describe the phenomenon of forming contact and clotting. Therefore, we focus on the contact intensity $\lambda_c$, the expected number of contacts per fiber $\lambda_{cF}$, and the expected number of clots per fiber $\lambda_{clF} = \frac{\lambda_{cl}}{\lambda_F}$, where $\lambda_{cl}$ is the expected number of clots per unit volume.
For algorithms whose runtime depends on the volume, the decision on the chosen size and the corresponding number, or vice versa, to achieve an RVE is a tradeoff depending on the runtime. In the following, we will refer to this as (RVE) size-number combination. To make the decision based on runtime considerations, we will not only carry out an RVE estimation following Kanit et~al.~\cite{kanit2003DeterminationSizeRepresentative}, but we will also report the observed runtime to justify our chosen size-number combination. More precisely, we carry it out on the 'most average' case of our experimental setup (see Section~\ref{sec:setup}), namely medium fiber aspect ratio $a = 30$, volume fraction $V_V = 20\%$, isotropic orientation ($\beta = 1.0$), and a curvature of $\kappa_1 = 10, \kappa_2 = 100$ under the interaction distance $\varepsilon = 0.3$. We generate 50\,realizations each for the window edge lengths $s = 64, 128, 256, 384, 512, 640$. For each parameter combination, we estimate the expected number of contacts per fiber.\footnote {The contact intensity $\lambda_c$ is, in this case, only a scaled version of the expected number of contacts per fiber $\lambda_{cF}$; therefore, we omit it here.} $\lambda_{cF}$, and the  expected number of clots per fiber $\lambda_{clF}$. For each property $Z = \lambda_{cF}, \lambda_{clF}$, we fit a model of the variance $D_Z^2$ of the volume $V$ as
\begin{equation}
    D_Z^2(V) = KV^{-\alpha}, ~~~ K, \alpha > 0,
\end{equation}
based on the mean value $\bar{Z}$ of characteristic $Z$ on the volume $V$; see also Dirrenberger et~al.~\cite{dirrenberger2014GiganticRVESizes}. The factor $K^\frac{1}{3} $ is a multiple of the generalized integral range, which can be interpreted as the scale of the phenomenon~\cite{kanit2003DeterminationSizeRepresentative}, which in our case is forming contact and clotting.
 For a relative precision $\varphi$ for the estimation of $Z$, we can then estimate the required number of simulations when using volume V to get overall representative information, short RVE number as~\cite{dirrenberger2014GiganticRVESizes}:
\begin{equation}
    n_{\text{RVE}}(Z, V) = \frac{4}{\varphi \bar{Z}}\frac{K}{V^{\alpha}}.
\end{equation}

\begin{figure}[ht!]
    \centering
    \includegraphics[width=0.6\textwidth]{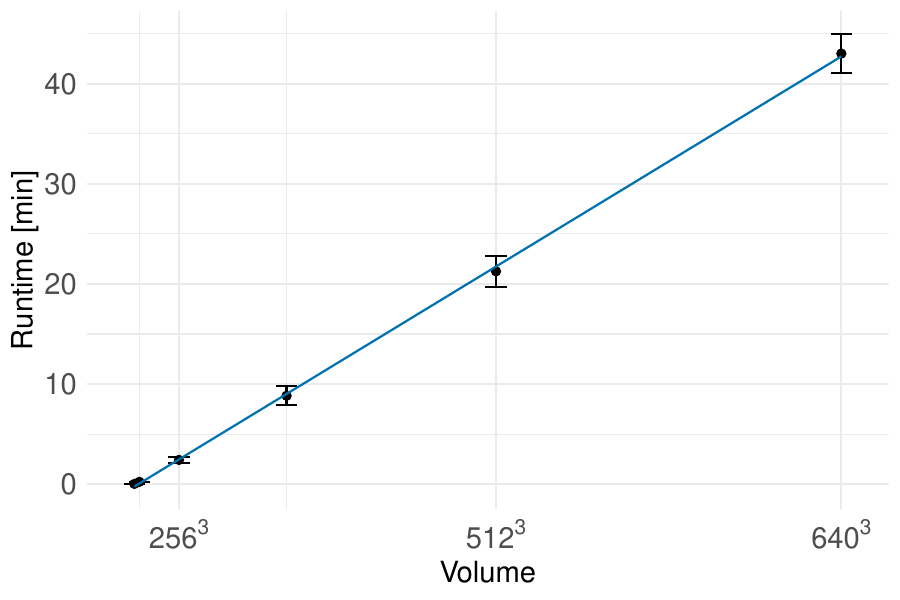}
    \caption{Plot for the runtime depending on the window size for medium fiber length and volume fraction, $\beta = 1.0$ and the curvature $\kappa_1 = 10, \kappa_2 = 100$. The scale of the window size is transformed cubically (corresponding to the volume). Bars indicate the standard deviation, the line in blue indicates a fitted model.}
    \label{fig:runtime}
\end{figure}

The runtime of the contact packing is proportional to the volume, see Fig.~\ref{fig:runtime}. It has a runtime of roughly 10\,min for size 384, which is acceptable when considering 'many' parameters in a simulation study, as we do in this section~\ref{sec:validation}. Note that this runtime increases with the volume fraction, or rather, the number of fibers in a window of constant size.

For each characteristic $Z$ and window size $s$, we fit a model of the variance $D_Z^2$ to the simulation results and report the RVE number in Table~\ref{table:RVEnumber} using a relative precision of $1\%$.  For the expected number of contacts per fiber $\lambda_{cF}$, we found $K_{cF}^\frac{1}{3} \simeq 51.65$ and $\alpha_{cF} \simeq -0.9954$. For the the expected number of clots per fiber $\lambda_{clF}$, we found $K_{clF}^\frac{1}{3} \simeq 14.91$ and $\alpha_{clF} \simeq -1.0020$. Notably, both $K_{cF}^\frac{1}{3}$ and $K_{cl}^\frac{1}{3}$ are smaller than the fiber length of $120$. Plots of $D_Z^2$ w.r.t. the volume and the fitted model can be found in Fig.~\ref{fig:RVEEstimation} in the Appendix.

\begin{table}[ht!]
\caption{The number of realizations necessary for a given size and characteristic for a relative precision of $1\%$.}
\label{table:RVEnumber}
\centering
\begin{tabular}{|r||r|r|r|} 
 \hline
 Size & $\lambda_{cF}$ & $\lambda_{cl}$ \\
 \hline\hline
 64 & 420 & 797\\
 128 & 46 & 87\\
 256 & 6 & 11 \\
 384 & 2 & 4\\
 512 & 1 & 2\\
 640 & 1 & 1\\
 \hline
\end{tabular}
\end{table}

In the following, we will generate 4 realizations on a window size of 384 for each parameter combination: As Table~\ref{table:RVEnumber} shows, only $4$ realizations of window size $384$ are necessary to achieve a relative precision of $1\%$. This yields the lowest runtime, see Fig.~\ref{fig:runtime}, in comparison to the other RVE size-number combinations, while having a larger window size than the longest fiber length $\ell = 200$.

\subsection{Contact Intensity Achieved Depending on the Interaction Distance}
\label{sec:intensityVsDistance}
\begin{figure}[ht!]
    \centering
    \begin{subfigure}{0.95\textwidth}
        \includegraphics[width=0.99\textwidth]{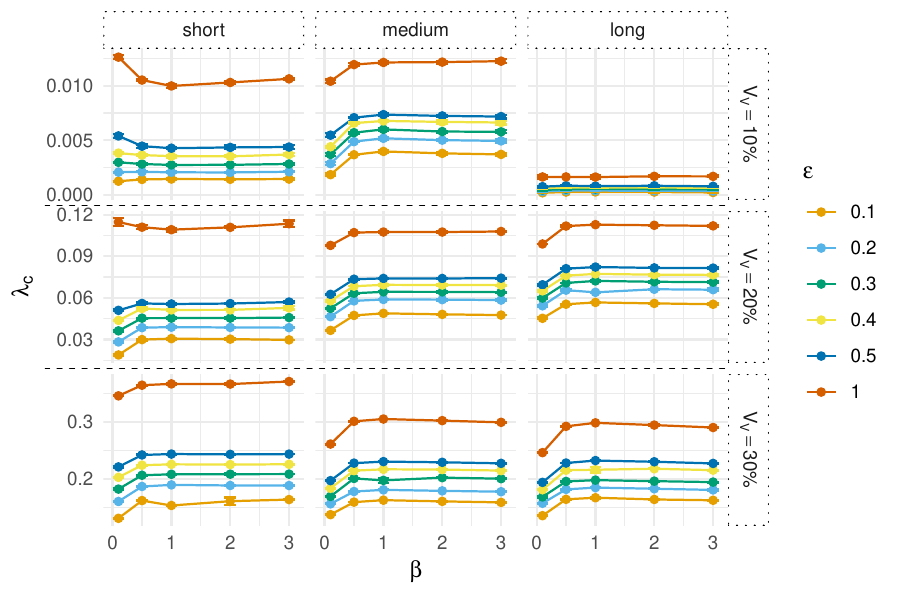}
    \end{subfigure}
    \caption{Plot of the contact intensity $\lambda_c$ w.r.t. the parameter of the orientation distribution for varying interaction distance $\varepsilon$ (in colour) for the experimental setup described in Section~\ref{sec:setup}. Note that $\kappa_1 = 10, \kappa_2 = 100$.}
    \label{fig:contactDensityVsInteractionDistance}
\end{figure}

The contact packing does not contain the contact intensity as a direct input parameter; instead, it contains the interaction distance, which does not directly translate to the contact intensity. Therefore, we investigate the relationship between the interaction distance and contact intensity in this subsection. Results are given in Fig.~\ref{fig:contactDensityVsInteractionDistance}.

\begin{figure}[ht!]
    \centering
    \begin{subfigure}{0.95\textwidth}
        \includegraphics[width=0.99\textwidth]{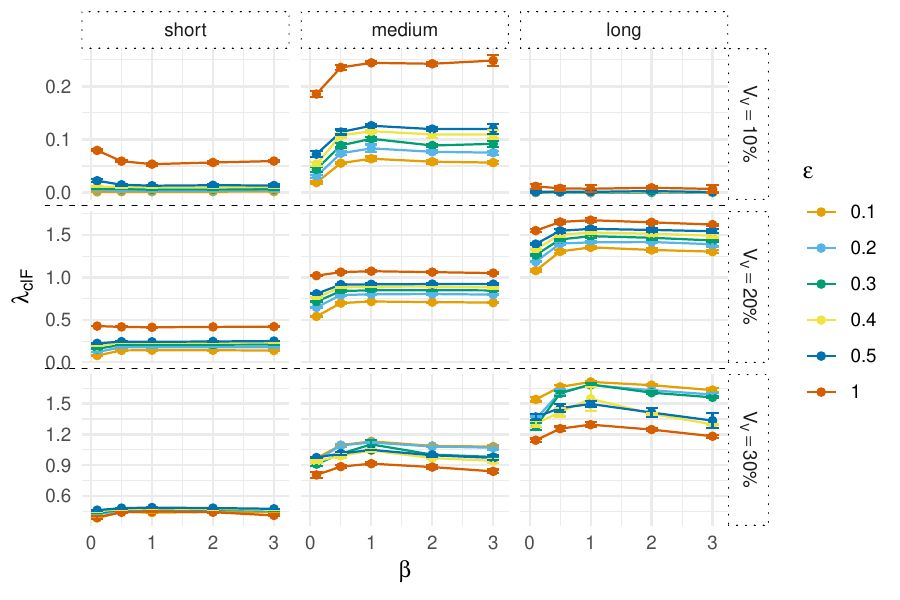}
    \end{subfigure}
    \caption{Plot of the clot density $\lambda_{cl}$ w.r.t. the parameter of the orientation distribution for varying interaction distance $\varepsilon$ (in colour) for the experimental setup described in Section~\ref{sec:setup}. Note that $\kappa_1 = 10, \kappa_2 = 100$.}
    \label{fig:clotDensity}
\end{figure}

The contact intensity is not quite linear w.r.t. the interaction distance, but Fig.~\ref{fig:contactDensityVsInteractionDistance} provides an indication of what to expect for similar setups. For each parameter combination, the contact intensity is relatively constant for $\beta \geq 1.0$ but generally smaller for $\beta < 1.0$. A likely reason for this is that for isotropic and girdle distributions, i.e., $\beta \geq 1.0$, there are more balls of different fibers in the neighborhood of each ball. For more aligned distributions, i.e., $\beta < 1.0$, the neighboring balls are more likely to belong to the same fiber, which yields larger contact areas instead. The same trend is observed for the Toll estimate, which increase with $\beta$ as $-\beta\log\beta$ when $\beta$ is small (see Eq.~\ref{eq:betafinal}, Appendix~\ref{sec:Poiss}).
Validating this conjecture will be part of future work. 
Notably, this effect reverses for low volume fractions, short fibers, and high interaction distances. Since the fibers are smaller, the likelihood that balls belong to different fibers increases again. This translates to more fibers clustering, as is shown in Fig.~\ref{fig:clotDensity}.

For long fibers and low volume fraction, the chosen interaction distance barely achieves an increase in contact intensity. This is most likely due to the distance between single fibers. 
This also conforms with the high parameter accuracy in this case, see Section~\ref{sec:accuracy}. Similarly, curvy fibers result in slightly lower contact and clot densities than their straighter counterparts, see Fig.~\ref{fig:densitiesStraight} and~\ref{fig:densitiesCurvy}.

\subsection{Achieved Contact Intensity Compared to Toll's Formula}
In order to compare the contact intensity between both packing algorithms and Toll's formula~\cite{toll1998PackingMechanicsFiber}, see Section~\ref{sec:analyticalModels}, we study the realizations for each parameter combination proposed in the section above with the interaction distance $\varepsilon = 1.0$. We estimate their contact intensity $\hat\lambda_c$, see Section~\ref{sec:analyticalModels}, both after the Altendorf-Jeulin packing and the contact packing.

\begin{figure}[h!]
    \centering
    \begin{subfigure}{0.95\textwidth}
        \includegraphics[width=0.99\textwidth]{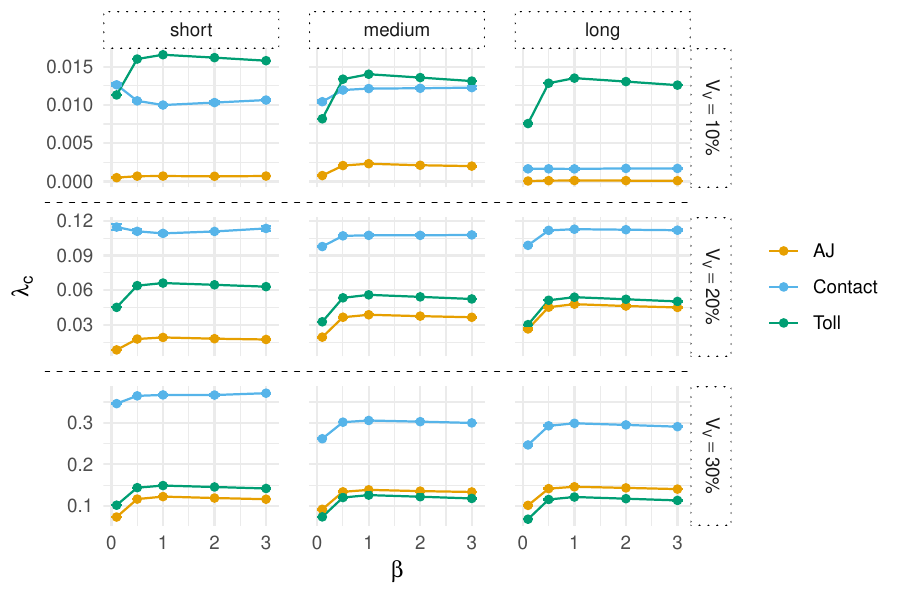}
    \end{subfigure}
    \caption{Plot of the contact intensity of the Altendorf-Jeulin model, the contact model, and the analytically determined formula by Toll. Note that $\kappa_1 = 10, \kappa_2 = 100$.}
    \label{fig:contactDensity}
\end{figure}

Fig.~\ref{fig:contactDensity} shows the results for the intensities. The already low contact intensities for the low volume fraction mostly stay below Toll's intersection intensity. This may be improved by omitting the prepacking in the Altendorf-Jeulin model. Yet for most other cases, the contact intensity can be raised above Toll's intensity. For high volume fraction and at least medium aspect ratio, the contact intensity even exceeds Toll's intensity after the Altendorf-Jeulin packing. This effect is not surprising considering that hardcore packing excludes a considerable volume for fibers, whereas the Boolean model does not have this constraint.

\begin{figure}[h!]
    \centering
        \includegraphics[width=0.8\textwidth]{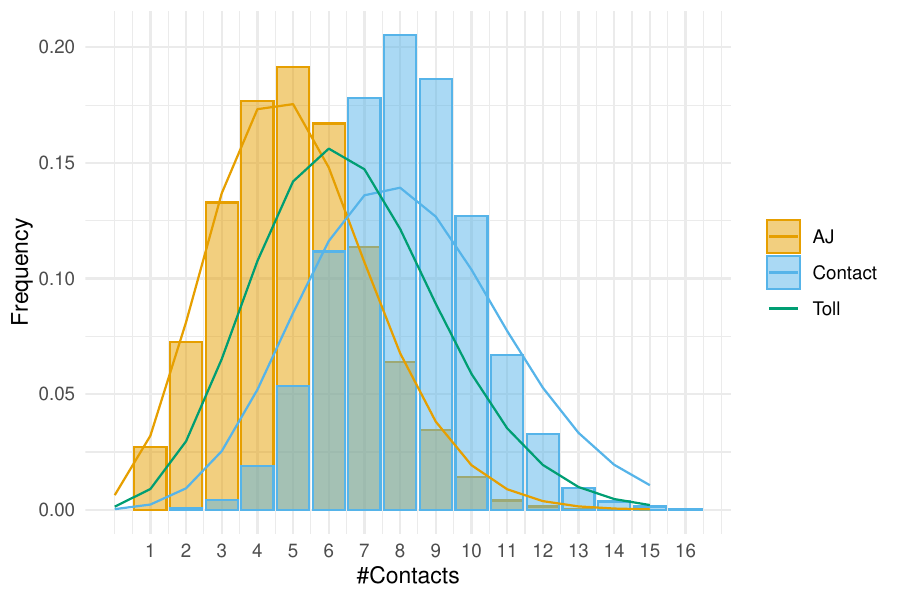}
    \caption{Plot of the distribution of the number of contacts of individual fibers for the Altendorf-Jeulin model and the contact model (colored bars). Solid lines represent Poisson distributions 
    with the same mean as that of the two models (left and right)
    as well as that predicted by the Toll model (in-between). Note that $\kappa_1 = 10, \kappa_2 = 100$.}
    \label{fig:contactStats}
\end{figure}

The probability distribution for the
number of contacts per fiber
is represented in Fig.~\ref{fig:contactStats} for the medium aspect ratio $a = 30$,
a volume fraction $V_V = 20 \%$, an isotropic orientation distribution, and an interaction distance of $\varepsilon
= 1.0$. 
Results obtained for the Altendorf-Jeulin and our contact
models are compared with the
distribution obtained for the Toll model~(\ref{eq:toll}).
This distribution is obtained in a similar way as the
mean number of contact points and follows a Poisson law
in the case of an isotropic distribution of orientation
(see Appendix~\ref{sec:Poiss} for details).
Furthermore, we fit the numerical results 
obtained for the Altendorf-Jeulin and contact models
with Poisson laws.

The Toll model gives a Poisson distribution with a mean of $6.6$ contacts per fiber. In the AJ model, it has a mean of roughly $5$ contact points and is very close to a Poisson distribution. In the contact model, the mean is about $8.2$, which is much higher.
Furthermore, the contact distribution is more narrow than in a Poisson distribution. Note, however, that the Toll model gives relevant estimates for intersecting fibers.
The absence of interpenetations in the contact model
is a considerable constraint that induce in particular repulsion effects and anti-correlations not taken into account in the Toll model.

\newpage
\subsection{Accuracy Results}
\label{sec:accuracy}
In order to evaluate the accuracy of our packing scheme, we estimate the direction parameter $\hat{\beta}$ and the curvature parameters $\hat{\kappa}_1$ and $\hat{\kappa}_2$ as proposed and implemented by Altendorf~\cite{altendorf2011RandomwalkbasedStochasticModeling}. We do this both after fiber generation but before packing, after packing with the Altendorf-Jeulin method, and after contact packing with interaction distance $\varepsilon = 0.1$.
We report their relative deviation from the original input parameters as an indicator of accuracy. Note that $0$ indicates high accuracy due to no deviation, and $1$ indicates low accuracy due to high deviation. Being particularly interested in any possible inaccuracy introduced by the contact packing steps, we also compare the deviation between the Altendorf-Jeulin packing and the contact packing.

\begin{figure}[H]
    \centering
    \begin{subfigure}{0.95\textwidth}
        \includegraphics[width=0.99\textwidth]{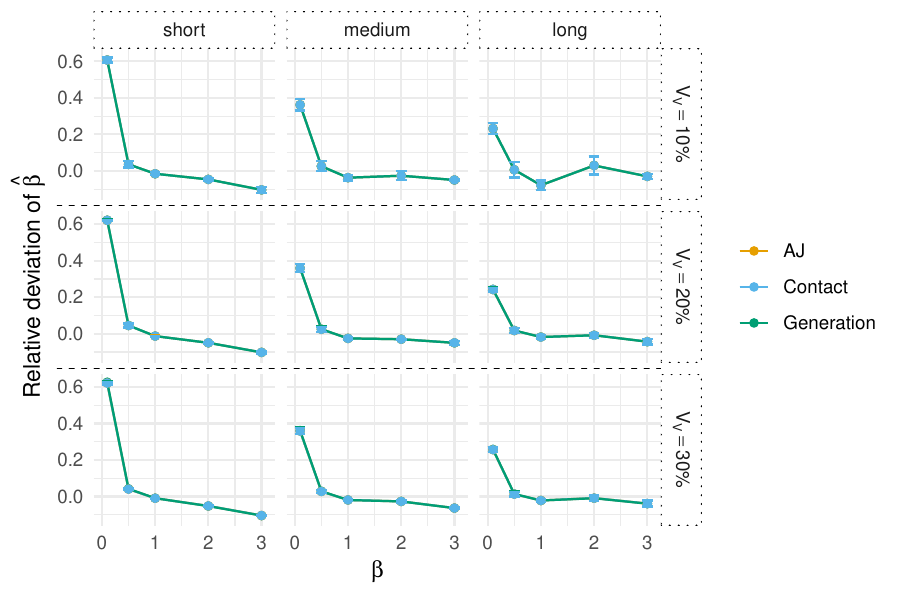}
    \end{subfigure}
    \caption{Plots of the deviation from the direction parameter $\beta$ relative to the input value. The estimated values of $\beta$ for the contact packing and the Altendorf-Jeulin packing correspond so closely to the values upon fiber generation that the plots overlay each other. Bars indicate standard deviations, but are often not visible because they are so low. Note that low deviation around $0$ indicates high accuracy, whereas high values indicate low accuracy. Note that $\kappa_1 = 10, \kappa_2 = 100$.}
    \label{fig:accuracyBeta}
\end{figure}

\begin{figure}[H]
    \centering
        \begin{subfigure}{0.95\textwidth}
        \includegraphics[width=0.99\textwidth]{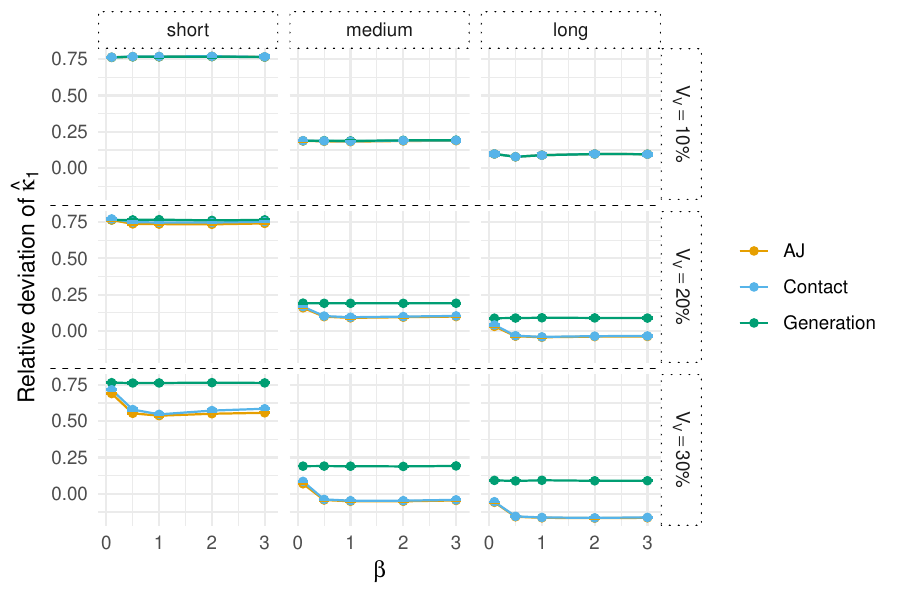}
    \end{subfigure}
    \begin{subfigure}{0.95\textwidth}
        \includegraphics[width=0.99\textwidth]{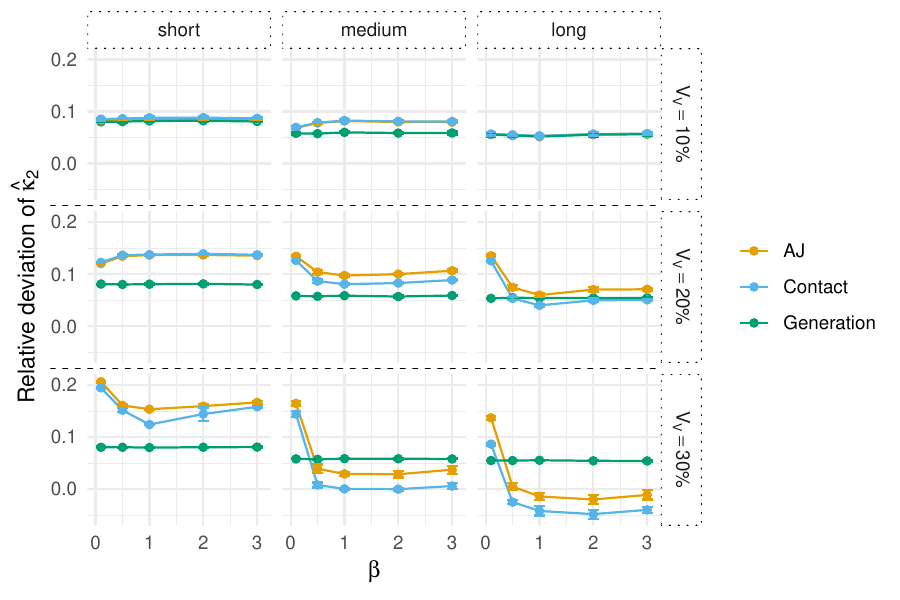}
    \end{subfigure}
    \caption{Plots of the deviation of the curvature parameters $\kappa_1 = 10$ and $\kappa_2 = 100$ relative to the input values after fiber generation, Altendorf-Jeulin packing, and contact packing each. Bars indicate standard deviations, but are often not visible because they are so low. Note that low values for the absolute deviation around $0$ indicate high accuracy, whereas high values indicate low accuracy.
    \label{fig:accuracyKappa}}
\end{figure}

All methods (fiber generation, Altendorf-Jeulin packing, contact packing) show acceptable accuracy for the fiber direction parameter $\beta$ for values of $\beta \geq 0.5$, see Fig.~\ref{fig:accuracyBeta}: The relative deviation from the input parameter is below $0.1$. For smaller values, however, the relative deviation rises up to $0.6$, even more so for shorter fibers. The estimated value of the curvature parameters $\kappa_1$ and $\kappa_2$, however, shows rather high inaccuracy: Whereas $\kappa_2$ has a relative deviation from the input parameter below just $0.2$, the relative deviation can even rise up to even $0.75$ for the parameter $\kappa_1$, see Fig.~\ref{fig:accuracyKappa}. This is especially pronounced for shorter fibers. Note that it is already alluded to in previous works that the curvature parameters are qualitatively meaningful, but their quantitative estimation is "not highly accurate"~\cite{altendorf20113DMorphologicalAnalysis}.

\begin{figure}[h!]
    \centering
    \begin{subfigure}{0.95\textwidth}
        \includegraphics[width=0.99\textwidth]{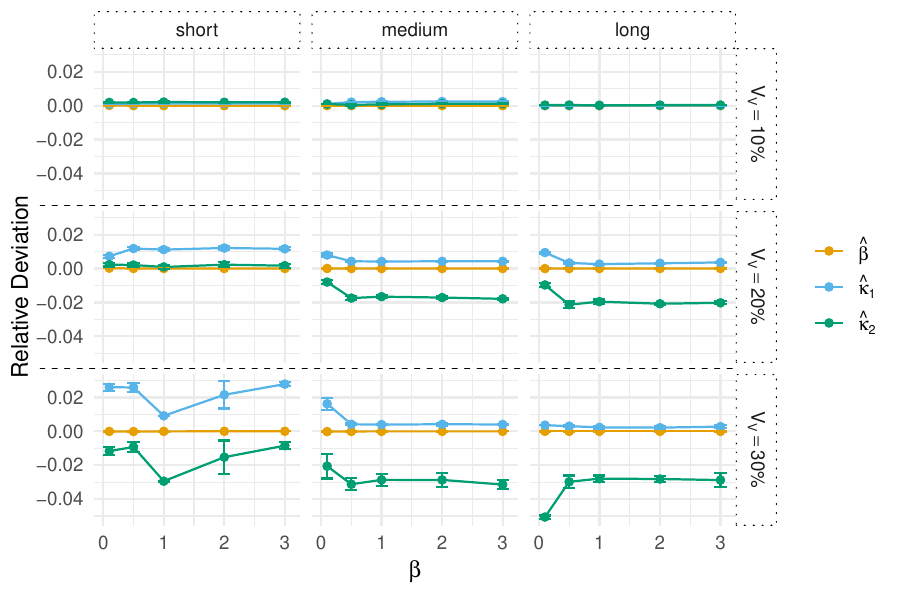}
    \end{subfigure}
    \caption{Deviation of the estimated parameter values of the contact model relative to the Altendorf-Jeulin model. Bars indicate standard deviations, but are often not visible because they are so low. Note that low values for the absolute deviation around $0$ indicate high accuracy, whereas high values indicate low accuracy. In this setup, the curvature parameters are $\kappa_1 = 10, \kappa_2 = 100$.}
    \label{fig:relDeviation}
\end{figure}

Fig.~\ref{fig:relDeviation} shows the relative deviation of the contact packing to the Altendorf-Jeulin packing. Whereas the direction parameter $\beta$ shows only very small deviation, i.e., it stays highly consistent, one can observe higher deviation for the local curvature $\kappa_2$. This is an intuitive result considering that the contact packing creates contact by attracting balls of fibers that are close to each other, which likely increases local curvature (corresponding to lower values). Nevertheless, the deviation of the curvature should be taken with a grain of salt, given the inaccuracy of their estimation observed above.

\newpage


\newpage

\newpage

\section{Application}
\label{sec:application}
In the present paper, we study the contact intensity of fiber models due to their role in materials' physical properties. So far, we have solely studied the models' behavior depending on input parameters. In this section, we study its behavior when applied to a real-world problem, namely, modeling thermal conductivity in wood fiber insulation mats. We will replicate the models generated by Andr\"a et~al.~\cite{andra2023ImagebasedMicrostructuralSimulation}. However, we will use the Altendorf-Jeulin and the contact model instead of a Boolean model. Note that simulating the thermal conductivity is a part of future work.

We replicate  Andr\"a et~al.'s base model M1 on a voxel spacing of 25\,\textmu m with the following parameter combinations: The mean fiber radius is $r = 50$\,\textmu m, the fiber length $\ell = 1$\,mm, the orientation parameter $\beta = 3.0$, indicating a fiber distribution closely aligned with the x-y-plane, and the curvature parameters $\kappa_1, \kappa_2 = 100$. We generate 6\,480 fibers to generate a volume fraction of roughly 6\,\%. This accounts for the lumen, the tunnel in wood fibers, modeled by Andr\"a et~al.~\cite{andra2023ImagebasedMicrostructuralSimulation}. Note that we omit the fiber chunks here, thus yielding slightly changed volume fractions.

As proposed by Andr\"a et~al.~\cite{andra2023ImagebasedMicrostructuralSimulation}, we vary the radius as $r = 35$\,\textmu m, 45\,\textmu m, 55\,\textmu m, 65\,\textmu m, the length as $\ell = 0.5$\,mm, 2\,mm, 3\,mm, the orientation parameter as $\beta = 3.5, 4.0$, each such that we keep all other parameters constant, and the number of fibers as 2\,160, 4\,320, 8\,640, 10\,800 to yield volume fractions of roughly $V_V = 2\%, 4\%, 8\%, 10\%$. For each combination, we plot the contact intensity for the Altendorf-Jeulin model, the contact model for interaction distance $\varepsilon = 1.0$, and Toll's intersection intensity~\cite{toll1993NoteTubeModel, toll1998PackingMechanicsFiber} for the Boolean model.

\begin{figure}[ht!]
    \centering
    \begin{subfigure}{0.49\textwidth}
        \includegraphics[width=0.99\textwidth]{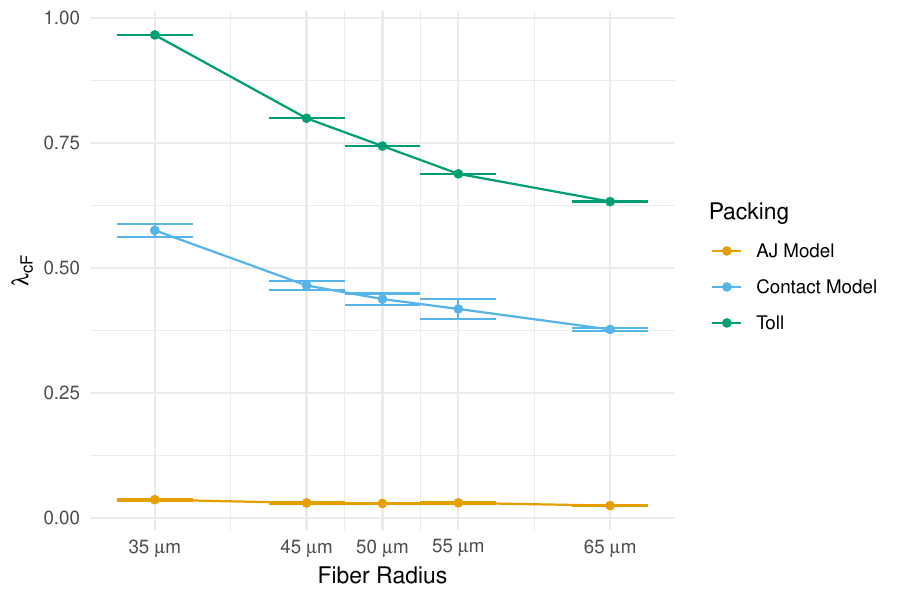}
        \caption{Contact density for varying radius.}
    \end{subfigure}
    \begin{subfigure}{0.49\textwidth}
        \includegraphics[width=0.99\textwidth]{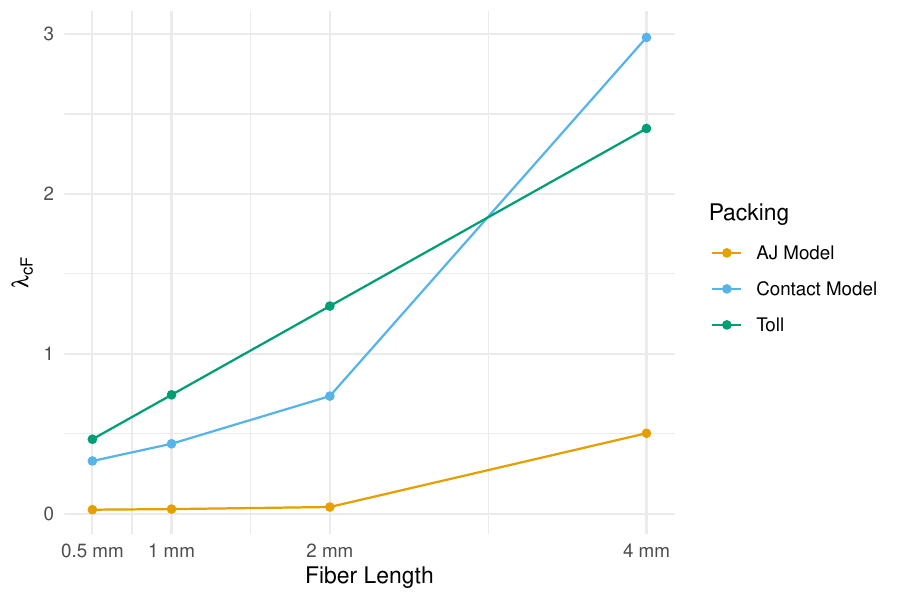}
        \caption{Contact density for varying length.}
    \end{subfigure}
    \begin{subfigure}{0.49\textwidth}
        \includegraphics[width=0.99\textwidth]{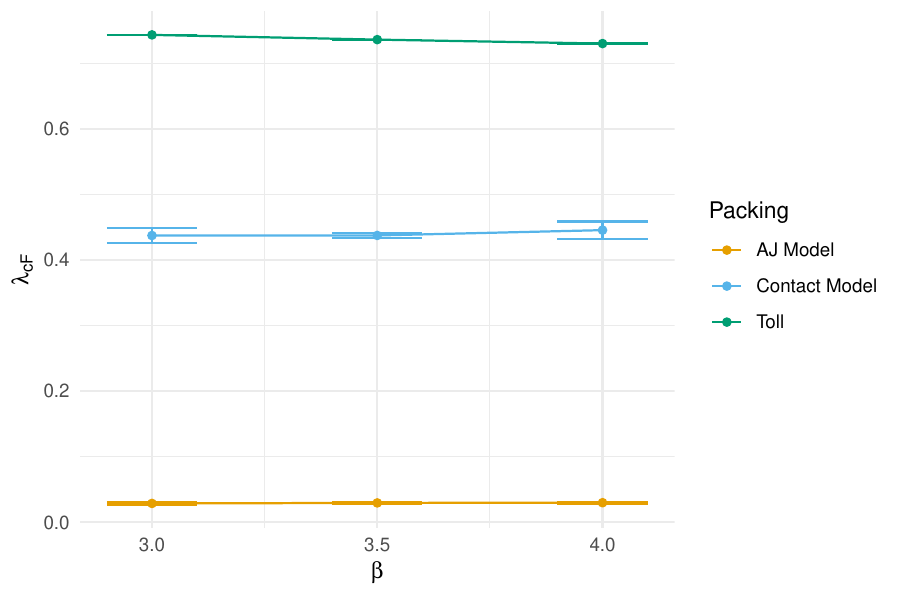}
        \caption{Contact density for varying orientation parameter $\beta$.}
    \end{subfigure}
    \begin{subfigure}{0.49\textwidth}
        \includegraphics[width=0.99\textwidth]{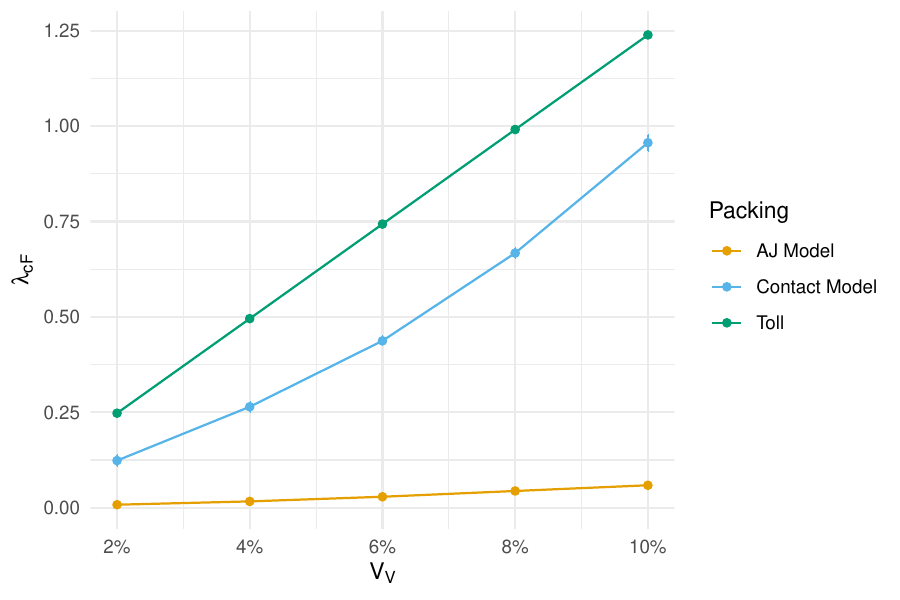}
        \caption{Contact density for varying volume fraction.}
    \end{subfigure}
    \caption{Contact densities when replicating the models proposed by Andr\"a et~al.~\cite{andra2023ImagebasedMicrostructuralSimulation}. Note that $\kappa_1, \kappa_2 = 100$.}
    \label{fig:application}
\end{figure}

We observe from Fig.~\ref{fig:application} that increasing the fiber radius reduces the contact density, whereas increasing the fiber length increases the density. This is not surprising as Toll's formula is directly dependent on the aspect ratio of fibers and the volume fraction. Interestingly, though, the contact model's density seems to be a fraction of Toll's density, except for the case of high fiber length. Here, the density of the contact model even rises higher than Toll's density.

The results indicate that modelling the fiber system with a Boolean model or a hardcore model may result in strongly differing conductivities due to the difference in contact densities. Moreover, the impact of the number of contacts in comparison to other parameters can be studied by systematic variation, which is planned for future work.
\section{Discussion \& Conclusion}
\label{sec:conclusion}
Despite the relevance of inter-fiber contacts when simulating physical properties of fiber systems, there exists little research on parametrically modelling fiber systems including inter-fiber contact. Therefore, we extend the force-biased model by Altendorf \& Jeulin~\cite{altendorf2011RandomwalkbasedStochasticModeling} to a model where the contact between fibers can be explicitly increased. For this, we have developed algorithmic notions of contact and closeness. To increase the number of contacts, we have added another force to the model by Altendorf \& Jeulin that creates contact between fibers that are close enough.

We have shown that our model indeed increases the number of contacts in comparison to the Altendorf-Jeulin model while maintaining similar accuracy to other parameters of the fiber system. Moreover, we compared the number of contacts in the algorithmic hardcore models with the intersection intensity of the Boolean model, finding that the contact intensity of the contact model exceeds the intersection intensity of the Boolean model for volume fractions above 20\,\%.

When simulating physical properties on fiber systems realized from the contact model, this allows for higher accuracy regarding the intensity of inter-fiber contacts. Moreover, the contact intensity can be systematically varied, thus allowing for a deeper understanding of the influence of contact on the physical property in comparison to other parameters. Such a study for thermal conductivity in wood fiber mats is currently underway. So far, the contact intensity can be reliably influenced by the interaction distance. To allow for an explicit parameter in the fiber packing algorithm for the contact intensity, further research is needed.

Studying the accuracy of the curvature was severely limited since, to our knowledge, no accurate estimators for the parameters $\kappa_1$ and $\kappa_2$ exist~\cite{altendorf2011RandomwalkbasedStochasticModeling}. In some cases, it may suffice to use only one curvature parameter, thus allowing for higher accuracy.

In the next stage, we will extend the fiber model to contain fiber bundles and potentially different shapes such as square cross sections. Moreover, studying and varying the contact surface in addition to the number (intensity) of contacts is worthwhile to gain a deeper understanding of the influence of contact on physical properties.

\section*{Acknowledgements}
We thank Markus Kronenberger and Michael Godehardt, Fraunhofer Institute for Industrial Mathematics (ITWM), for their support with the MAVIlib library and access to the ITWM Beehive Cluster.
This work was supported by the German Federal Ministry of Education and Research under Grant Agreement No:~05M22UKA and the French-German doctoral college "Mathematische Bildverarbeitung". 
\section*{Code and Data Availability}
The code for the contact model can be provided upon reasonable request. The result data and corresponding scripts for evaluation are hosted on \url{doi:10.5281/zenodo.17105506}.

\section*{Author Contributions}
\textbf{Alex Keilmann: } Data curation, Formal analysis, Methodology, Software, Validation, Visualization, Writing - original draft, Writing - review \& editing. \textbf{Claudia Redenbach: } Conceptualization, Writing - review \& editing . \textbf{Fran\c{c}ois Willot: }  Conceptualization, Writing - review \& editing.

\bibliographystyle{myunsrtnat.bst}  
\bibliography{biblio, biblioExtern, biblioFW}

\begin{thebibliography}{51}
\providecommand{\natexlab}[1]{#1}
\providecommand{\url}[1]{\texttt{#1}}
\expandafter\ifx\csname urlstyle\endcsname\relax
  \providecommand{\doi}[1]{doi: #1}\else
  \providecommand{\doi}{doi: \begingroup \urlstyle{rm}\Url}\fi

\bibitem[Andr{\"a} et~al.(2023)Andr{\"a}, Dobrovolskij, Engelhardt, Godehardt, Makas, Mercier, Rief, Schladitz, Staub, Trawka, and Treml]{andra2023ImagebasedMicrostructuralSimulation}
H. Andr{\"a}, D. Dobrovolskij, M. Engelhardt, M. Godehardt, M. Makas, C. Mercier, S. Rief, K. Schladitz, S. Staub, K. Trawka, and S. Treml.
\newblock Image-based microstructural simulation of thermal conductivity for highly porous wood fiber insulation boards: {{3D}} imaging, microstructure modeling, and numerical simulations for insight into structure--property relation.
\newblock \emph{Wood Science and Technology}, 57\penalty0 (1):\penalty0 5--31, 2023.
\newblock \doi{10.1007/s00226-022-01434-6}.

\bibitem[Berhan and Sastry(2007)]{berhan07Modeling}
L. Berhan and A.M. Sastry.
\newblock Modeling percolation in high-aspect-ratio fiber systems. i. soft-core versus hard-core models.
\newblock \emph{Phys. Rev. E}, 75:\penalty0 041120, 2007.
\newblock \doi{10.1103/PhysRevE.75.041120}.

\bibitem[Pan et~al.(2008)Pan, Iorga, and Pelegri]{pan2008Analysis3DRandom}
Y. Pan, L. Iorga, and A.A. Pelegri.
\newblock Analysis of {{3D}} random chopped fiber reinforced composites using {{FEM}} and random sequential adsorption.
\newblock \emph{Computational Materials Science}, 43\penalty0 (3):\penalty0 450--461, 2008.
\newblock \doi{10.1016/j.commatsci.2007.12.016}.

\bibitem[Redenbach and Vecchio(2011)]{redenbach2011StatisticalAnalysisStochastic}
C. Redenbach and I. Vecchio.
\newblock Statistical analysis and stochastic modelling of fibre composites.
\newblock \emph{Composites Science and Technology}, 71\penalty0 (2):\penalty0 107--112, 2011.
\newblock \doi{10.1016/j.compscitech.2010.10.014}.

\bibitem[Tian et~al.(2015)Tian, Qi, Zhou, Liang, and Ma]{tian2015RepresentativeVolumeElement}
W. Tian, L. Qi, J. Zhou, J. Liang, and Y. Ma.
\newblock Representative volume element for composites reinforced by spatially randomly distributed discontinuous fibers and its applications.
\newblock \emph{Composite Structures}, 131:\penalty0 366--373, 2015.
\newblock \doi{10.1016/j.compstruct.2015.05.014}.

\bibitem[Provatas et~al.(2000)Provatas, Haataja, Asikainen, Majaniemi, Alava, and {Ala-Nissila}]{provatas2000FiberDepositionModels}
N. Provatas, M. Haataja, J. Asikainen, S. Majaniemi, M. Alava, and T. {Ala-Nissila}.
\newblock Fiber deposition models in two and three spatial dimensions.
\newblock \emph{Colloids and Surfaces A: Physicochemical and Engineering Aspects}, 165\penalty0 (1-3):\penalty0 209--229, 2000.
\newblock \doi{10.1016/S0927-7757(99)00417-3}.

\bibitem[Niskanen and Alava(1994)]{niskanen1994PlanarRandomNetworks}
K.J. Niskanen and M.J. Alava.
\newblock Planar {{Random Networks}} with {{Flexible Fibers}}.
\newblock \emph{Physical Review Letters}, 73\penalty0 (25):\penalty0 3475--3478, 1994.
\newblock \doi{10.1103/PhysRevLett.73.3475}.

\bibitem[Venkateshan et~al.(2016)Venkateshan, Tahir, Vahedi~Tafreshi, and Pourdeyhimi]{venkateshan2016ModelingEffectsFiber}
D. Venkateshan, M. Tahir, H. Vahedi~Tafreshi, and B. Pourdeyhimi.
\newblock Modeling effects of fiber rigidity on thickness and porosity of virtual electrospun mats.
\newblock \emph{Materials \& Design}, 96:\penalty0 27--35, 2016.
\newblock \doi{10.1016/j.matdes.2016.01.105}.

\bibitem[Moghadam et~al.(2019)Moghadam, Yousefi, Tafreshi, and Pourdeyhimi]{moghadam2019CharacterizingNonwovenMaterials}
A. Moghadam, S. Yousefi, H.V. Tafreshi, and B. Pourdeyhimi.
\newblock Characterizing nonwoven materials via realistic microstructural modeling.
\newblock \emph{Separation and Purification Technology}, 211:\penalty0 602--609, 2019.
\newblock \doi{10.1016/j.seppur.2018.10.018}.

\bibitem[Williams and Philipse(2003)]{williams2003RandomPackingsSpheres}
S.R. Williams and A.P. Philipse.
\newblock Random packings of spheres and spherocylinders simulated by mechanical contraction.
\newblock \emph{Physical Review E}, 67\penalty0 (5):\penalty0 051301, 2003.
\newblock \doi{10.1103/PhysRevE.67.051301}.

\bibitem[Karayiannis and Laso(2008)]{karayiannis2008DenseNearlyJammed}
N.C. Karayiannis and M. Laso.
\newblock Dense and {{Nearly Jammed Random Packings}} of {{Freely Jointed Chains}} of {{Tangent Hard Spheres}}.
\newblock \emph{Physical Review Letters}, 100\penalty0 (5):\penalty0 050602, 2008.
\newblock \doi{10.1103/PhysRevLett.100.050602}.

\bibitem[Schneider(2017)]{schneider2017SequentialAdditionMigration}
M. Schneider.
\newblock The sequential addition and migration method to generate representative volume elements for the homogenization of short fiber reinforced plastics.
\newblock \emph{Computational Mechanics}, 59\penalty0 (2):\penalty0 247--263, 2017.
\newblock \doi{10.1007/s00466-016-1350-7}.

\bibitem[Schneider(2022)]{schneider2022AlgorithmGeneratingMicrostructures}
M. Schneider.
\newblock An algorithm for generating microstructures of fiber-reinforced composites with long fibers.
\newblock \emph{International Journal for Numerical Methods in Engineering}, 123\penalty0 (24):\penalty0 6197--6219, 2022.
\newblock \doi{10.1002/nme.7110}.

\bibitem[Mehta and Schneider(2022)]{mehta2022SequentialAdditionMigration}
A. Mehta and M. Schneider.
\newblock A sequential addition and migration method for generating microstructures of short fibers with prescribed length distribution.
\newblock \emph{Computational Mechanics}, 70\penalty0 (4):\penalty0 829--851, 2022.
\newblock \doi{10.1007/s00466-022-02201-x}.

\bibitem[Lauff et~al.(2024)Lauff, Schneider, Montesano, and B{\"o}hlke]{lauff2024GeneratingMicrostructuresLong}
C. Lauff, M. Schneider, J. Montesano, and T. B{\"o}hlke.
\newblock Generating microstructures of long fiber reinforced composites by the fused sequential addition and migration method.
\newblock \emph{International Journal for Numerical Methods in Engineering}, page e7573, 2024.
\newblock \doi{10.1002/nme.7573}.

\bibitem[Lauff et~al.(2025{\natexlab{a}})Lauff, Schneider, and B{\"o}hlke]{lauff2025MicrostructureGenerationLong}
C. Lauff, M. Schneider, and T. B{\"o}hlke.
\newblock Microstructure generation of long fiber reinforced hybrid composites using the fused sequential addition and migration method.
\newblock \emph{Journal of Thermoplastic Composite Materials}, page 08927057251314425, 2025{\natexlab{a}}.
\newblock \doi{10.1177/08927057251314425}.

\bibitem[Lauff et~al.(2025{\natexlab{b}})Lauff, Krause, Schneider, and B{\"o}hlke]{lauff2025InfluenceFiberCurvature}
C. Lauff, M. Krause, M. Schneider, and T. B{\"o}hlke.
\newblock On the {{Influence}} of the {{Fiber Curvature}} on the {{Stiffness}} of {{Long Fiber Reinforced Composites}}.
\newblock \emph{International Journal for Numerical Methods in Engineering}, 126\penalty0 (15):\penalty0 e70094, 2025{\natexlab{b}}.
\newblock \doi{10.1002/nme.70094}.

\bibitem[Klar(2019)]{klar19InteractingFiberStructures}
A. Klar.
\newblock Interacting fiber structures: mathematical aspects and applications.
\newblock \emph{Rivista di Matematica della Università di Parma}, 10\penalty0 (2):\penalty0 199--268, 2019.

\bibitem[Borsche et~al.(2017)Borsche, Klar, Nessler, Roth, and Tse]{borsche2017RetardedMeanFieldApproach}
R. Borsche, A. Klar, C. Nessler, A. Roth, and O. Tse.
\newblock A {{Retarded Mean-Field Approach}} for {{Interacting Fiber Structures}}.
\newblock \emph{Multiscale Modeling \& Simulation}, 15\penalty0 (3):\penalty0 1130--1154, 2017.
\newblock \doi{10.1137/151005592}.

\bibitem[Salnikov et~al.(2015)Salnikov, Cho{\"i}, and {Karamian-Surville}]{salnikov2015EfficientReliableStochastic}
V. Salnikov, D. Cho{\"i}, and P. {Karamian-Surville}.
\newblock On efficient and reliable stochastic generation of {{RVEs}} for analysis of composites within the framework of homogenization.
\newblock \emph{Computational Mechanics}, 55\penalty0 (1):\penalty0 127--144, 2015.
\newblock \doi{10.1007/s00466-014-1086-1}.

\bibitem[Bezrukov and Stoyan(2006)]{bezrukov2006SimulationStatisticalAnalysis}
A. Bezrukov and D. Stoyan.
\newblock Simulation and {{Statistical Analysis}} of {{Random Packings}} of {{Ellipsoids}}.
\newblock \emph{Particle \& Particle Systems Characterization}, 23\penalty0 (5):\penalty0 388--398, 2006.
\newblock \doi{10.1002/ppsc.200600974}.

\bibitem[Altendorf and Jeulin(2011)]{altendorf2011RandomwalkbasedStochasticModeling}
H. Altendorf and D. Jeulin.
\newblock Random-walk-based stochastic modeling of three-dimensional fiber systems.
\newblock \emph{Physical Review E}, 83\penalty0 (4):\penalty0 041804, 2011.
\newblock \doi{10.1103/PhysRevE.83.041804}.

\bibitem[Easwaran(2017)]{easwaran2017StochasticGeometryModels}
P. Easwaran.
\newblock \emph{Stochastic Geometry Models for Interacting Fibers}.
\newblock PhD thesis, TU Kaiserslautern, 2017.

\bibitem[Chapelle et~al.(2015)Chapelle, L{\'e}vesque, Br{\o}ndsted, Foldschack, and Kusano]{chapelle2015GENERATIONNONOVERLAPPINGFIBER}
L. Chapelle, M. L{\'e}vesque, P. Br{\o}ndsted, M.R. Foldschack, and Y. Kusano.
\newblock {{GENERATION OF NON-OVERLAPPING FIBER ARCHITECTURE}}.
\newblock \emph{Proceedings of the 20th {I}nternational {C}onference on {C}omposite {M}aterials}, 2015.

\bibitem[Kumar et~al.(2024)Kumar, DasGupta, and Jain]{kumar2024MicrostructureGenerationAlgorithm}
A. Kumar, A. DasGupta, and A. Jain.
\newblock Microstructure generation algorithm and micromechanics of curved fiber composites with random waviness.
\newblock \emph{International Journal of Solids and Structures}, 289:\penalty0 112625, 2024.
\newblock \doi{10.1016/j.ijsolstr.2023.112625}.

\bibitem[Gaunand et~al.(2025)Gaunand, De~Wilde, Fran\ifmmode~\mbox{\c{c}}\else \c{c}\fi{}ois, Grigorova-Moutiers, and Joulain]{Gaunand07Modeling}
C. Gaunand, Y. De~Wilde, A. Fran\ifmmode~\mbox{\c{c}}\else \c{c}\fi{}ois, V. Grigorova-Moutiers, and K. Joulain.
\newblock Modeling conductive thermal transport in three-dimensional fibrous media with fiber-to-fiber contacts.
\newblock \emph{Phys. Rev. Appl.}, 23:\penalty0 034084, 2025.
\newblock \doi{10.1103/PhysRevApplied.23.034084}.

\bibitem[Yastrebov(2013)]{yastrebov2013numerical}
V.A. Yastrebov.
\newblock \emph{Numerical methods in contact mechanics}.
\newblock John Wiley \& Sons, 2013.

\bibitem[Picu(2011)]{picu2011mechanics}
R. Picu.
\newblock Mechanics of random fiber networks—a review.
\newblock \emph{Soft Matter}, 7\penalty0 (15):\penalty0 6768--6785, 2011.

\bibitem[Xie et~al.(2023)Xie, Guo, Shao, Zhu, Jiao, Yang, and Chen]{xie2023mechanics}
J. Xie, Z. Guo, M. Shao, W. Zhu, W. Jiao, Z. Yang, and L. Chen.
\newblock Mechanics of textiles used as composite preforms: A review.
\newblock \emph{Composite structures}, 304:\penalty0 116401, 2023.

\bibitem[Poquillon et~al.(2005)Poquillon, Viguier, and Andrieu]{poquillon2005experimental}
D. Poquillon, B. Viguier, and E. Andrieu.
\newblock Experimental data about mechanical behaviour during compression tests for various matted fibres.
\newblock \emph{Journal of materials science}, 40:\penalty0 5963--5970, 2005.

\bibitem[Van~Wyk(1946)]{van194620}
C. Van~Wyk.
\newblock 20—note on the compressibility of wool.
\newblock \emph{Journal of the Textile Institute Transactions}, 37\penalty0 (12):\penalty0 T285--T292, 1946.

\bibitem[Durville(2005)]{durville2005numerical}
D. Durville.
\newblock Numerical simulation of entangled materials mechanical properties.
\newblock \emph{Journal of materials science}, 40:\penalty0 5941--5948, 2005.

\bibitem[Faessel et~al.(2005)Faessel, Delis{\'e}e, Bos, and Cast{\'e}ra]{faessel20053DModellingRandom}
M. Faessel, C. Delis{\'e}e, F. Bos, and P. Cast{\'e}ra.
\newblock {{3D Modelling}} of random cellulosic fibrous networks based on {{X-ray}} tomography and image analysis.
\newblock \emph{Composites Science and Technology}, 65\penalty0 (13):\penalty0 1931--1940, 2005.
\newblock \doi{10.1016/j.compscitech.2004.12.038}.

\bibitem[Karako{\c c} et~al.(2017)Karako{\c c}, Hiltunen, and Paltakari]{karakoc2017GeometricalSpatialEffects}
A. Karako{\c c}, E. Hiltunen, and J. Paltakari.
\newblock Geometrical and spatial effects on fiber network connectivity.
\newblock \emph{Composite Structures}, 168:\penalty0 335--344, 2017.
\newblock \doi{10.1016/j.compstruct.2017.02.062}.

\bibitem[Deogekar et~al.(2019)Deogekar, Yan, and Picu]{deogekar2019RandomFiberNetworks}
S. Deogekar, Z. Yan, and R.C. Picu.
\newblock Random {{Fiber Networks With Superior Properties Through Network Topology Control}}.
\newblock \emph{Journal of Applied Mechanics}, 86\penalty0 (8):\penalty0 081010, 2019.
\newblock \doi{10.1115/1.4043828}.

\bibitem[Huisman et~al.(2008)Huisman, Storm, and Barkema]{huisman2008MonteCarloStudy}
E.M. Huisman, C. Storm, and G.T. Barkema.
\newblock Monte {{Carlo}} study of multiply crosslinked semiflexible polymer networks.
\newblock \emph{Physical Review E}, 78\penalty0 (5):\penalty0 051801, 2008.
\newblock \doi{10.1103/PhysRevE.78.051801}.

\bibitem[Huisman et~al.(2007)Huisman, Van~Dillen, Onck, and Van Der~Giessen]{huisman2007ThreeDimensionalCrossLinkedFActin}
E.M. Huisman, T. Van~Dillen, P.R. Onck, and E. Van Der~Giessen.
\newblock Three-{{Dimensional Cross-Linked F-Actin Networks}}: {{Relation}} between {{Network Architecture}} and {{Mechanical Behavior}}.
\newblock \emph{Physical Review Letters}, 99\penalty0 (20):\penalty0 208103, 2007.
\newblock \doi{10.1103/PhysRevLett.99.208103}.

\bibitem[Toll(1993)]{toll1993NoteTubeModel}
S. Toll.
\newblock Note: {{On}} the tube model for fiber suspensions.
\newblock \emph{Journal of Rheology}, 37\penalty0 (1):\penalty0 123--125, 1993.
\newblock \doi{10.1122/1.550460}.

\bibitem[Toll(1998)]{toll1998PackingMechanicsFiber}
S. Toll.
\newblock Packing mechanics of fiber reinforcements.
\newblock \emph{Polymer Engineering \& Science}, 38\penalty0 (8):\penalty0 1337--1350, 1998.
\newblock \doi{10.1002/pen.10304}.

\bibitem[Komori and Makishima(1977)]{komori1977NumbersFiberFiberContacts}
T. Komori and K. Makishima.
\newblock Numbers of {{Fiber-to-Fiber Contacts}} in {{General Fiber Assemblies}}.
\newblock \emph{Textile Research Journal}, 47\penalty0 (1):\penalty0 13--17, 1977.
\newblock \doi{10.1177/004051757704700104}.

\bibitem[Gaiselmann et~al.(2013)Gaiselmann, Froning, T{\"o}tzke, Quick, Manke, Lehnert, and Schmidt]{gaiselmann2013Stochastic3DModeling}
G. Gaiselmann, D. Froning, C. T{\"o}tzke, C. Quick, I. Manke, W. Lehnert, and V. Schmidt.
\newblock Stochastic {{3D}} modeling of non-woven materials with wet-proofing agent.
\newblock \emph{International Journal of Hydrogen Energy}, 38\penalty0 (20):\penalty0 8448--8460, 2013.
\newblock \doi{10.1016/j.ijhydene.2013.04.144}.

\bibitem[Schladitz et~al.(2006)Schladitz, Peters, {Reinel-Bitzer}, Wiegmann, and Ohser]{schladitz2006DesignAcousticTrim}
K. Schladitz, S. Peters, D. {Reinel-Bitzer}, A. Wiegmann, and J. Ohser.
\newblock Design of acoustic trim based on geometric modeling and flow simulation for non-woven.
\newblock \emph{Computational Materials Science}, 38\penalty0 (1):\penalty0 56--66, 2006.
\newblock \doi{10.1016/j.commatsci.2006.01.018}.

\bibitem[Fisher et~al.(1987)Fisher, Lewis, and Embleton]{fisher1987StatisticalAnalysisSpherical}
N.I. Fisher, T. Lewis, and B.J.J. Embleton.
\newblock \emph{Statistical Analysis of Spherical Data}.
\newblock Cambridge University Press, Cambridge [Cambridgeshire] ; New York, 1987.
\newblock ISBN 978-0-521-24273-8.

\bibitem[Altendorf(2011)]{altendorf20113DMorphologicalAnalysis}
H. Altendorf.
\newblock \emph{{{3D Morphological Analysis}} and {{Modeling}} of {{Random Fiber Networks}}}.
\newblock PhD thesis, TU Kaiserslautern, Kaiserslautern, Germany, 2011.

\bibitem[Benoit et~al.(2002)Benoit, Corraze, and Chauvet]{Benoit_2002}
J.M. Benoit, B. Corraze, and O. Chauvet.
\newblock Localization, coulomb interactions, and electrical heating in single-wall carbon nanotubes/polymer composites.
\newblock \emph{Physical Review B}, 65\penalty0 (24), 2002.
\newblock \doi{10.1103/physrevb.65.241405}.

\bibitem[Mo{\'s}ci{\'n}ski et~al.(1989)Mo{\'s}ci{\'n}ski, Bargie{\l}, Rycerz, and Jacobs]{moscinski1989ForceBiasedAlgorithmIrregular}
J. Mo{\'s}ci{\'n}ski, M. Bargie{\l}, Z.A. Rycerz, and P.W.M. Jacobs.
\newblock The {{Force-Biased Algorithm}} for the {{Irregular Close Packing}} of {{Equal Hard Spheres}}.
\newblock \emph{Molecular Simulation}, 3\penalty0 (4):\penalty0 201--212, 1989.
\newblock \doi{10.1080/08927028908031373}.

\bibitem[Sedgewick and Wayne(2011)]{sedgewick2011algorithms}
R. Sedgewick and K. Wayne.
\newblock \emph{Algorithms}.
\newblock Addison-{W}esley {P}rofessional, 2011.

\bibitem[{Fraunhofer ITWM, Department of Image Processing}(2024)]{mavi}
{Fraunhofer ITWM, Department of Image Processing}.
\newblock {MAVI} -- modular algorithms for volume images, v. 2024.
\newblock \url{http://www.mavi-3d.de}, 2024.

\bibitem[Kanit et~al.(2003)Kanit, Forest, Galliet, Mounoury, and Jeulin]{kanit2003DeterminationSizeRepresentative}
T. Kanit, S. Forest, I. Galliet, V. Mounoury, and D. Jeulin.
\newblock Determination of the size of the representative volume element for random composites: Statistical and numerical approach.
\newblock \emph{International Journal of Solids and Structures}, 40\penalty0 (13-14):\penalty0 3647--3679, 2003.
\newblock \doi{10.1016/S0020-7683(03)00143-4}.

\bibitem[Dirrenberger et~al.(2014)Dirrenberger, Forest, and Jeulin]{dirrenberger2014GiganticRVESizes}
J. Dirrenberger, S. Forest, and D. Jeulin.
\newblock Towards gigantic {{RVE}} sizes for {{3D}} stochastic fibrous networks.
\newblock \emph{International Journal of Solids and Structures}, 51\penalty0 (2):\penalty0 359--376, 2014.
\newblock \doi{10.1016/j.ijsolstr.2013.10.011}.

\bibitem[Serra(1981)]{serra1981boolean}
J. Serra.
\newblock The boolean model and random sets.
\newblock In \emph{Image modeling}, pages 343--370. Elsevier, 1981.

\end{thebibliography}

\newpage

\appendix

\section{Parameter combinations in Section~\ref{sec:validation}}

In Section~\ref{sec:setup}, we explained the experimental setup for validating the contact model rather illustratively, omitting exact parameters. Here, in Table~\ref{table:parameterCombinations}, we provide detailed input parameters for the Altendorf-Jeulin and the contact model.
\begin{table}[!ht]
\caption{Overview of parameter combinations used in the experiments in Section~\ref{sec:validation}. All parameters were combined with $\beta = 0.1, 0.5, 1.0, 2.0, 3.0$ and $\varepsilon = 0.1, 0.2, 0.3, 0.4, 0.5, 1.0$. Note that the number of fibers depends on the image size~$s$.}
\label{table:parameterCombinations}
\centering
\begin{tabular}{|r|r|r|r|r|r|r|r|l|} 
 \hline
 Aspect ratio & Radius & Length & $V_V$~~~\, & $\#\text{fibers}$ & $\kappa_1$ & $\kappa_2$ & $\tau$~~ & $d_c$ \\
 \hline\hline
  10 & 2 &  40 & $\sim 10 \%$ &  $400 s^3$ & 10 & 100 & 0.0 & 0\\
  10 & 2 &  40 & $\sim 20 \%$ &  $800 s^3$ & 10 & 100 & 0.0 & 0\\
  10 & 2 &  40 & $\sim 30 \%$ &$1\,200 s^3$& 10 & 100 & 0.0 & 0\\
  30 & 2 & 120 & $\sim 10 \%$ &  $133 s^3$ & 10 & 100 & 0.0 & 0\\
  30 & 2 & 120 & $\sim 20 \%$ &  $266 s^3$ & 10 & 100 & 0.0 & 0\\
  30 & 2 & 120 & $\sim 30 \%$ &  $400 s^3$ & 10 & 100 & 0.0 & 0\\
  50 & 2 & 200 & $\sim 10 \%$ &   $80 s^3$ & 10 & 100 & 0.0 & 0\\
  50 & 2 & 200 & $\sim 20 \%$ &  $160 s^3$ & 10 & 100 & 0.0 & 0\\
  50 & 2 & 200 & $\sim 30 \%$ &  $240 s^3$ & 10 & 100 & 0.0 & 0\\
  30 & 2 & 120 & $\sim 20 \%$ &  $266 s^3$ & 10 &  10 & 0.0 & 0\\
  30 & 2 & 120 & $\sim 20 \%$ &  $266 s^3$ &100 & 100 & 0.0 & 0\\
 \hline
\end{tabular}
\end{table}

\section{Fitted Model for the Representative Volume Element}

\begin{figure}[ht!]
    \centering
    \begin{subfigure}{0.49\textwidth}
        \includegraphics[width=0.95\textwidth]{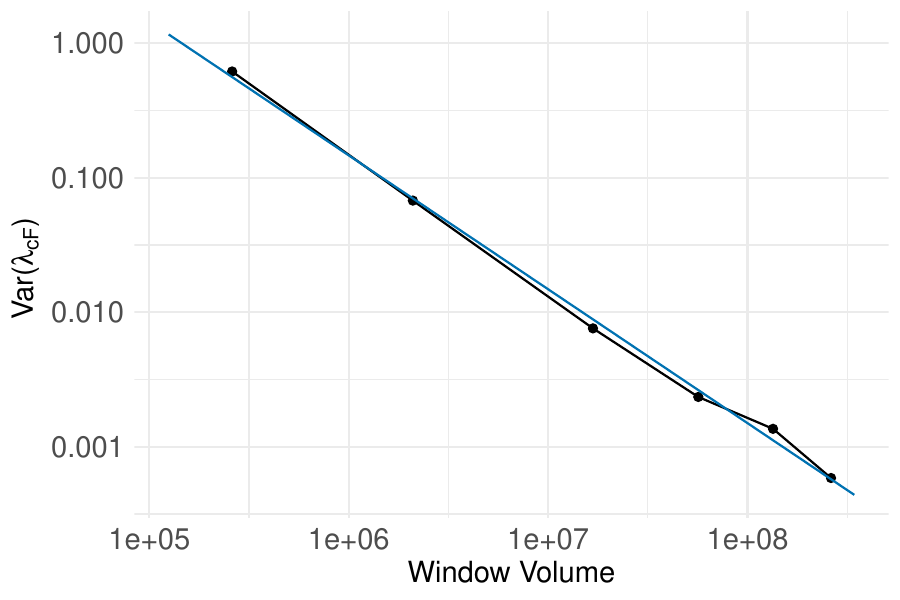}
    \end{subfigure}
    \begin{subfigure}{0.49\textwidth}
        \includegraphics[width=0.95\textwidth]{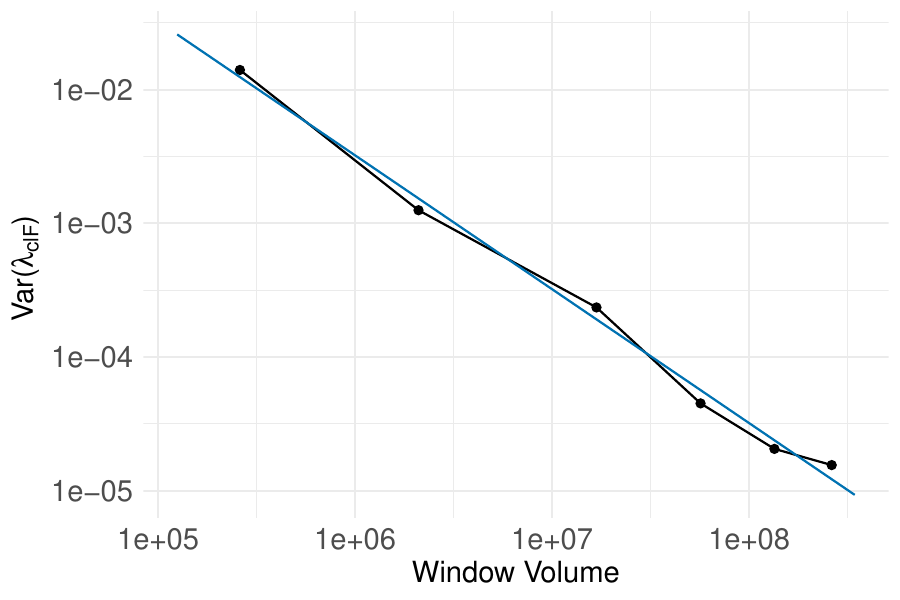}
    \end{subfigure}
    \caption{Plots of the variance of properties w.r.t. to the window volume. The line in {\color{oi_blue}blue} indicates the fitted model.}
    \label{fig:RVEEstimation}
\end{figure}

Fig.~\ref{fig:RVEEstimation} shows the variance plots w.r.t. the volume window, which is used in Section~\ref{sec:RVE}, together with their fitted models. These are used for the RVE size-number estimation. Note that we only present the plot for $Z = \lambda_{cF}, \lambda_{clF}$ as $\lambda_c$ is a multiple of $\lambda_{cF}$ in this case due to a constant $\lambda_F$. Note that the fitted models correspond closely to the observed variances.

\section{Validation Plots of Section~\ref{sec:validation}}

\begin{figure}[ht!]
    \centering
    \begin{subfigure}{0.49\textwidth}
        \includegraphics[width=0.99\textwidth]{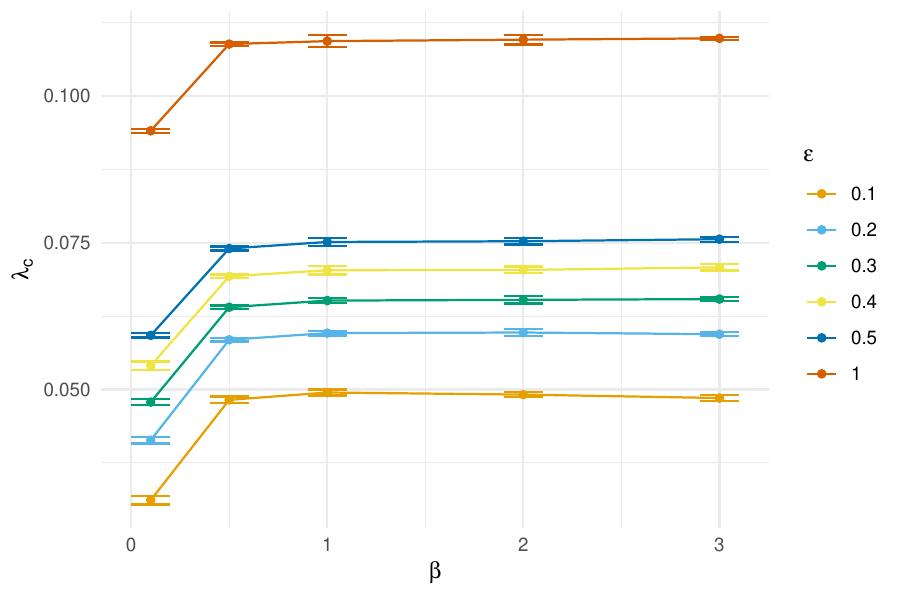}
    \end{subfigure}
    \begin{subfigure}{0.49\textwidth}
        \includegraphics[width=0.99\textwidth]{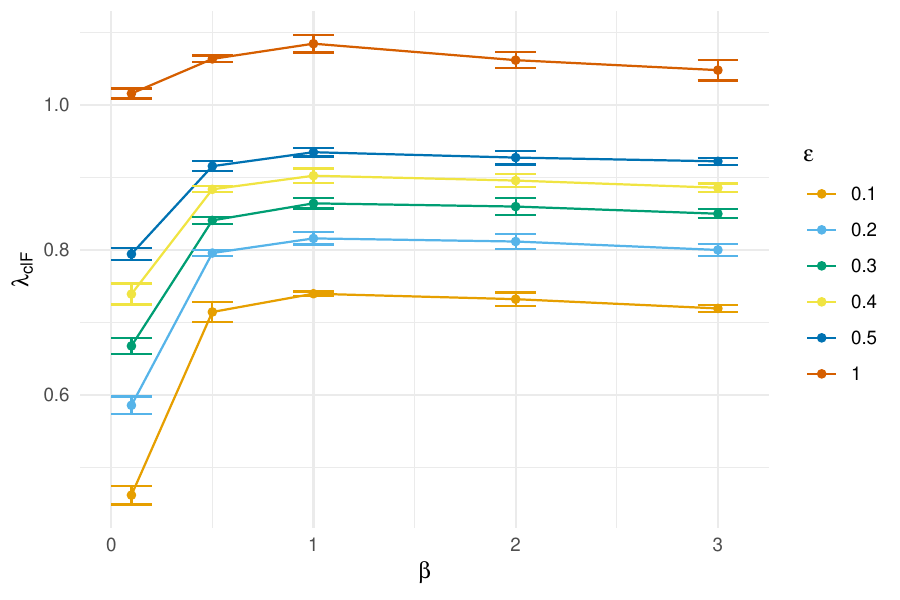}
    \end{subfigure}
    \caption{Plot of the contact intensity $\lambda_c$ and he expected number of clots per fiber $\lambda_{clF}$ w.r.t. the parameter of the orientation distribution for varying interaction distance $\varepsilon$ (in colour) for straight fibers with $\kappa_1, \kappa_2 = 100$, see Section~\ref{sec:intensityVsDistance}.}
    \label{fig:densitiesStraight}
\end{figure}

\begin{figure}[!ht]
    \centering
    \begin{subfigure}{0.49\textwidth}
        \includegraphics[width=0.99\textwidth]{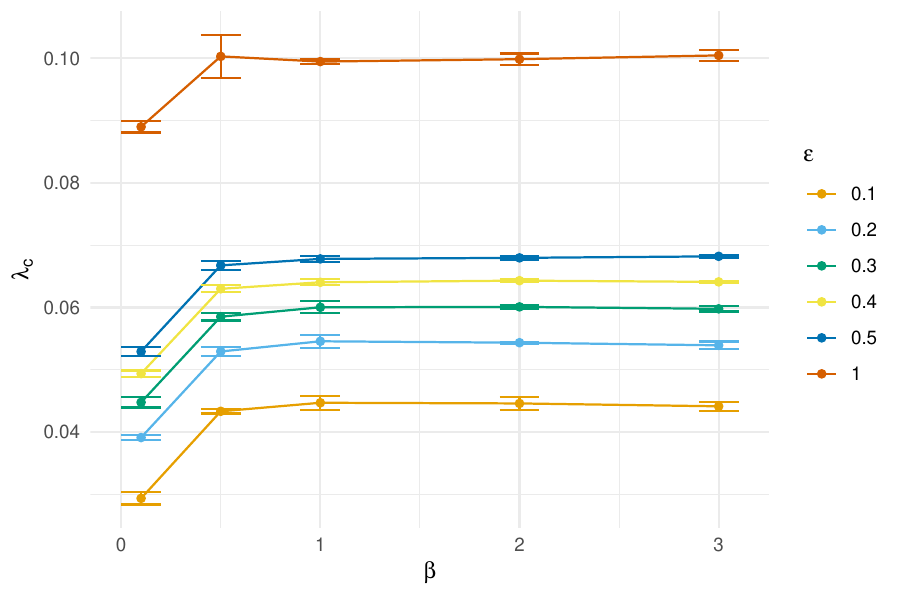}
    \end{subfigure}
    \begin{subfigure}{0.49\textwidth}
        \includegraphics[width=0.99\textwidth]{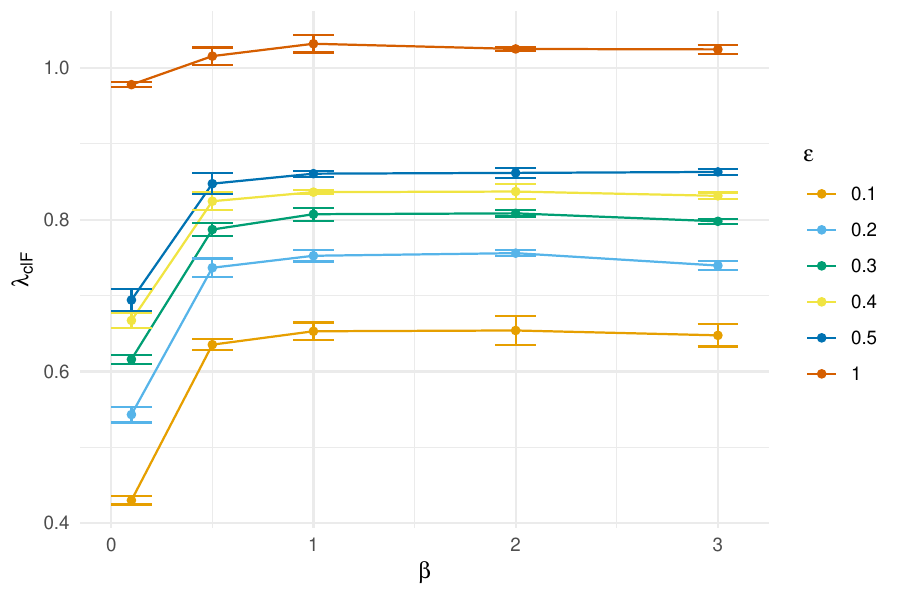}
    \end{subfigure}
    \caption{Plot of the contact intensity $\lambda_c$ and he expected number of clots per fiber $\lambda_{clF}$ w.r.t. the parameter of the orientation distribution for varying interaction distance $\varepsilon$ (in colour) for curvy fibers with $\kappa_1, \kappa_2 = 10$, see Section~\ref{sec:intensityVsDistance}.}
    \label{fig:densitiesCurvy}
\end{figure}

In Section~\ref{sec:validation}, the simulation results were shown for the main setup of moderate fiber curvature $\kappa_1 = 10, \kappa_2 = 100$. Fig.~\ref{fig:densitiesStraight} shows the impact of contact distance $\varepsilon$ on both the contact intensity $\lambda_c$ and the expected number of clots per fiber $\lambda_{clF}$ for straight fibers, so curvature parameters $\kappa_1, \kappa_2 = 100$. Fig.~\ref{fig:densitiesCurvy} again, shows these results for curvy fibers, meaning curvature parameters $\kappa_1, \kappa_2 = 10$. Their behavior is similar to fibers with moderate curvature $\kappa_1 = 10, \kappa_2 = 100$, yet differ in the specific values and observed variance of $\lambda_c$ and $\lambda_{clF}$.

\begin{figure}[ht!]
    \centering
    \begin{subfigure}{0.49\textwidth}
        \includegraphics[width=0.99\textwidth]{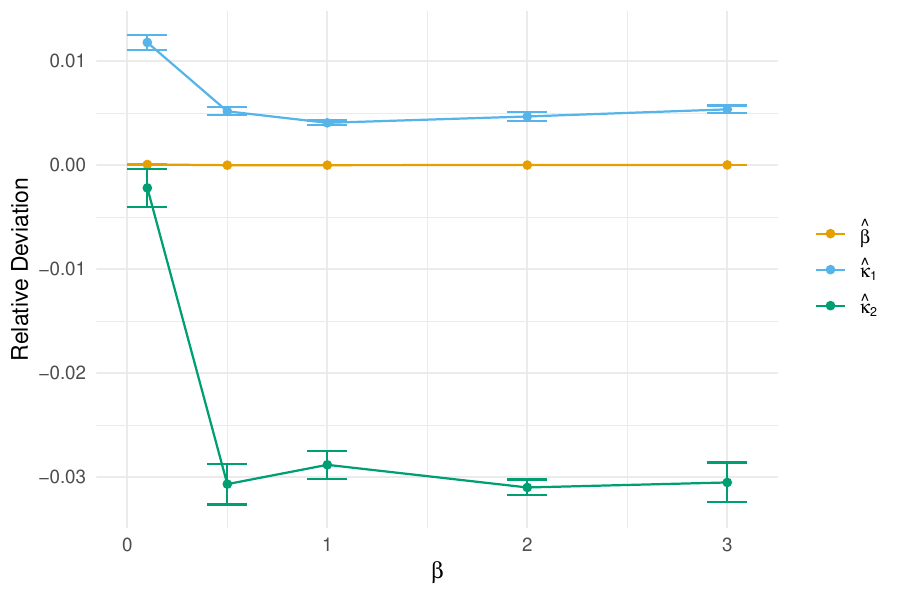}
        \caption{Relative deviations of straight fibers with $\kappa_1, \kappa_2 = 100$, see Section~\ref{sec:setup}.}
    \end{subfigure}
    \begin{subfigure}{0.49\textwidth}
        \includegraphics[width=0.99\textwidth]{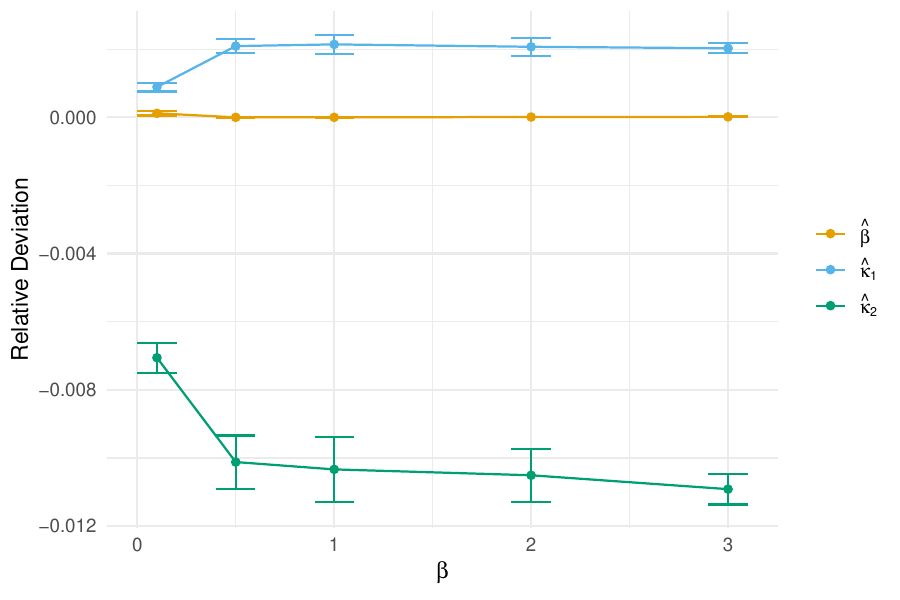}
        \caption{Relative deviations of curvy fibers, see Section~\ref{sec:setup}.}
    \end{subfigure}
    \caption{Deviation of the estimated parameter values of the contact model relative to the Altendorf-Jeulin model for the special cases of straight ($\kappa_1, \kappa_2 = 100$) and curvy ($\kappa_1, \kappa_2 = 10$) fibers. Bars indicate the standard deviation. Note that small values for the absolute deviation around $0$ indicate high accuracy, whereas high values indicate low accuracy.}
    \label{fig:relDeviationStraightCurvy}
\end{figure}

Fig.~\ref{fig:relDeviationStraightCurvy} presents relative deviation of estimated parameters for rather straight fibers ($\kappa_1, \kappa_2 = 100$) and more curvy fibers ($\kappa_1, \kappa_2 = 10$) relative to the corresponding Altendorf-Jeulin model. Note that the accuracy is higher for curvy fibers.

\section{Distribution of the number of fiber-fiber intersections}
\label{sec:Poiss}
Hereafter, we derive the distribution for the number of intersections of a test fiber with other fibers in a Boolean random set~\cite{serra1981boolean}.
The Boolean random set is made of cylinders
of radius $r$, length $\ell$,
volume $V=\pi r^2\ell$
and density $\lambda_f$.
The fibers are randomly oriented and follow the orientation distribution
$\psi(p)$ where $p$ is a unit vector
and:
\begin{equation}
    \int_{|p|=1} \psi(p){\rm d}p=1.
\end{equation}
Let us consider a cylinder oriented along direction $p$
and the centerline of an arbitrary cylinder oriented along direction $p'$.
We follow~\cite{toll1993NoteTubeModel} and note
that the \emph{centerline} will cut the test cylinder if and only if its center is inside a domain of volume:
\begin{equation}\label{eq:ss0}
A\ell + V,
\end{equation}
with $A$ the projected area of the test cylinder along a plane perpendicular to $p'$:
\begin{equation}\label{eq:ss0xxx}
A=2r\ell |p\times p'| + \pi r^2 |p\cdot p'|.
\end{equation}
The mean number of intersections between the test cylinder and the centerline of a cylinder oriented along direction $p$ therefore reads:
\begin{equation}\label{eq:errt}
\delta(p') \left( A\ell + V \right),
\end{equation}
where $\delta(p')=\lambda_f\psi(p'){\rm d}p'$
is the density of cylinders oriented
along direction $p'$.
Denote $\lambda'_i$ the number of intersects of a test cylinder
oriented along direction $p$
with a \emph{cylinder} oriented in direction $p'$.
Its mean value reads (see~\cite{toll1998PackingMechanicsFiber}):
\begin{equation}
\label{eq:ss0dde}
\overline{\lambda'_i}=
4\delta(p')\left(A'\ell + V\right), \qquad
A'=r\ell |p\times p'| + \pi r^2 |p\cdot p'|,
\end{equation}
and the variable $\lambda'_i$ follows the Poisson distribution:
\begin{equation}\label{eq:ss0b}
P\lbrace \lambda'_i=k \rbrace = 
\text{e}^{-\overline{\lambda'_i}}\frac{\overline{\lambda'_i}^k}{k!}, \qquad k\in\mathbb{N}.
\end{equation}
The total number of intersections
$\lambda_i$ between the test fiber and a fiber in any direction is
 the sum of the independent Poisson variables
$\lambda'_i$ and is therefore also Poisson-distributed.
Accordingly:
\begin{equation}\label{eq:ss0c}
P\lbrace \lambda_i=k \rbrace = 
\text{e}^{-\overline{\lambda_i}}\frac{\overline{\lambda_i}^k}{k!}, 
\qquad
\overline{\lambda_i}
=\int_{|p'|=1}
4\psi(p') \lambda_f(A'\ell+V){\rm d}p'.
\end{equation}
In the above, $\overline{\lambda_i}$ and $P\lbrace \lambda_i=k \rbrace$  depend, in general,
on the direction $p$ of the test cylinder.
Finally, the distribution of the number of
intersections $\widetilde{\lambda_i}$
of an arbitrarily-oriented cylinder with another cylinder is given by averaging~(\ref{eq:ss0c}) over all $p$:
\begin{equation}\label{eq:ss0cxxxq}
P\lbrace \widetilde{\lambda_i}=k \rbrace = 
\int_{p} {\rm d}p\, \psi(p)
\text{e}^{-\overline{\lambda_i}}\frac{\overline{\lambda_i}^k}{k!}, \qquad k\in\mathbb{N}.
\end{equation}
For a set of cylinders with the same length,
Eq.~(\ref{eq:toll}) can be deduced from the above, by taking the 
expectation of the probability law.
When fibers are isotropically-distributed ($\beta=1$) or aligned ($\beta=0$), 
Eq.~(\ref{eq:ss0cxxxq}) reduces to 
the same expression as in~(\ref{eq:ss0cxxxq})
and so $\widetilde{\lambda_i}$
is also Poisson-distributed.

Let us now focus on the particular case where $\psi(p)$
takes the form~(\ref{eq:betapdf})
and consider the limit $r\to\infty$ of infinitely-elongated cylinders.
The integral for $f_{\psi}$ in~(\ref{eq:toll}) reads, in polar coordinates:
\begin{equation}
f_{\psi}=
\int
_{[0;\pi]^2\times[0;2\pi]^2}
{\rm d}\theta\,{\rm d}\theta'\,{\rm d}\,\phi\,{\rm d}\phi'\,
\frac{\beta^2 \sin\theta\sin\theta'
\sqrt{1-[\sin\theta\sin\theta'\cos(\phi-\phi')+\cos\theta\cos\theta']^2}}{
4\pi^2
\left[(\beta^2-1)\cos^2\theta+1\right]^{3/2}
\left[(\beta^2-1)\cos^2(\theta')+1\right]^{3/2}
}.
\end{equation}
We denote $u=\phi-\phi'$ and expand the integrand as $\theta$, $\theta'\to 0$:
\begin{equation}
f_{\psi}\approx 
\int_{u=0}^{2\pi}
\int_{\theta=0}^{\pi/2}
\int_{\theta'=0}^{\pi/2}{\rm d}u\,{\rm d}\theta\,{\rm d}\theta'\,
\frac{u \beta^2 \theta\theta' \sqrt{\theta^2+\theta'^2-2\theta\theta' \cos u}}{
2\pi^2(\beta^2+\theta^2)^{3/2}(\beta^2+\theta'^2)^{3/2}},
\end{equation}
which provides, after a change of variable $w=\theta/\beta$, $w'=\theta'/\beta$
\begin{equation}
\frac{\partial (f_{\psi}/\beta)}{\partial w_{\max}} \approx
\int_{u=0}^{2\pi}{\rm d}u
\int_{w=1}^{w_{\max}}{\rm d}w
\frac{u \sqrt{w^2-2 w w_{\max} \cos u+w_{\max}^2}}{\pi^2 w^2 w_{\max}^2}
\approx
\int_{u=0}^{2\pi}{\rm d}u
\int_{w=1}^{w_{\max}}{\rm d}w
\frac{u}{\pi^2 w^2 w_{\max}}
\end{equation}
with $w_{\max}=\pi/(2\beta)$.
Finally, we obtain:
\begin{equation}\label{eq:b0}
f_{\psi}=-2\beta\log\beta+O(\beta), \qquad \beta\to 0,
\end{equation}
whereas $g_{\psi}\to 1$ (see~\cite{toll1998PackingMechanicsFiber}).
Therefore, in the limit 
of quasi-aligned fiebrs of infinite length, we obtain:
\begin{equation}\label{eq:betafinal}
\lambda_{cF}= 8 \lambda_{F} r\ell
(\pi r-\ell\beta\log\beta).
\end{equation}

\end{document}